\documentclass[usegraphicx]{mn2e}
\usepackage{subfigure}
\usepackage{epsf}
\usepackage{amsmath}

\newcommand {\apgt} {\ {\raise-.5ex\hbox{$\buildrel>\over\sim$}}\ }
\newcommand {\aplt} {\ {\raise-.5ex\hbox{$\buildrel<\over\sim$}}\ } 
\newcommand {\degree}{$^{\circ}$}

\title[A spectral-timing model for ULXs]
{A spectral-timing model for ULXs in the super-critical regime}

\author[M.Middleton et al.]
{Matthew J. Middleton$^{1}$, Lucy Heil$^{2}$, Fabio Pintore$^{3}$, Dominic J. Walton$^{4, 5}$ \newauthor and Timothy P. Roberts$^{6}$\\
\\
1. Institute of Astronomy, Madingly Road, Cambridge, UK\\
2. Astronomical Institute Anton Pannekoek, Science Park 904,1098 XH, Amsterdam, Netherlands\\
3. Department of Physics, Universit\`{a} di Cagliari, I-09042 Cagliari, Italy\\ 
4. NASA Jet Propulsion Laboratory, 4800 Oak Grove Dr, Pasadena, CA 91109, USA\\
5. Cahill Centre for Astronomy \& Astrophysics, California Institute of Technology, 1200 East California Boulevard,\\ Pasadena, California 91125, USA\\
6. Department of Physics, University of Durham, Durham DH1 3LE, UK
}

\pagerange{\pageref{firstpage}--\pageref{lastpage}} \pubyear{2011}
\long\def\symbolfootnote[#1]#2{\begingroup\def\thefootnote{\fnsymbol{footnote}}\footnote[#1]{#2}\endgroup} 
\def\ga{\mathrel{\hbox{\rlap{\hbox{\lower4pt\hbox{$\sim$}}}{\raise2pt\hbox{$>$}}
}}}

\begin{document}

\topmargin = -0.5cm

\maketitle

\label{firstpage}

\begin{abstract}

  Ultraluminous X-ray sources (ULXs) with luminosities lying between
  $\sim$3$\times$10$^{39}$-2$\times$10$^{40}$ erg s$^{-1}$ represent a
  contentious sample of objects as their brightness, together with a
  lack of unambiguous mass estimates for the vast majority of the central objects, leads to
  a degenerate scenario where the accretor could be a stellar remnant (black hole or neutron star) or
  intermediate mass black-hole (IMBH). Recent, high-quality
  observations imply that the presence of IMBHs in the
  {\it majority} of these objects is unlikely unless the accretion flow somehow deviates
  strongly from expectation based on objects with known masses. On the
  other-hand, physically motivated models for super-critical inflows can re-create the observed X-ray
  spectra and their evolution, although have been lacking a robust
  explanation for their variability properties. In this paper we
  include the effect of a partially inhomogeneous wind that imprints
  variability onto the X-ray emission via two distinct methods. The model is heavily dependent on both
  inclination to the line-of-sight and mass accretion rate, resulting
  in a series of qualitative and semi-quantitative predictions. We
  study the time-averaged 
  spectra and variability of a sample of well-observed ULXs,
  finding that the source behaviours can be explained by our model in both individual cases as well as across the entire sample, specifically in the trend of hardness-variability
  power. We present the covariance spectra for these sources for the first time, which shed light on the correlated variability and issues associated with modelling broad ULX spectra.

\end{abstract}

\section{Introduction}

Ultraluminous X-ray sources (ULXs) have been widely observed in the
local Universe, with inferred isotropic luminosities above 10$^{39}$
erg s$^{-1}$ (Roberts 2007; Feng \& Soria 2011). Those below
$\sim$3$\times$10$^{39}$ erg s$^{-1}$ can be readily associated with
accretion onto stellar mass black holes ($\sim$10 M$_{\odot}$)
accreting close to or at their Eddington limit (see Sutton, Roberts \&
Middleton 2013 and references therein). There is now strong evidence
to support this assertion, with the discovery of extremely bright
ballistic jets from a ULX in M31 (Middleton et al. 2013; Middleton, Miller-Jones \& Fender 2014), which
unambiguously links the flow with Eddington rate accretion (Fender,
Belloni \& Gallo 2004), and the first dynamical mass measurement of
the compact object in a ULX, from M101 ULX-1 (Liu et
al. 2013). Observations of such `low luminosity' ULXs (Soria et
al. 2012; Middleton et al. 2011a, 2012; Kaur et al. 2012) have
revealed changes in the disc emission that may imply the creation of a
radiation pressure supported, larger scale-height flow in the inner
regions (Middleton et al. 2012) or magnetic pressure support (Straub,
Done \& Middleton 2013). Although emission below $\sim$ 2~keV is
generally heavily photo-electrically absorbed by material in the
Galactic plane (e.g. Zimmerman et al. 2001), similar spectral
behaviour may also be seen in a small number of Galactic black hole
X-ray binaries (BHBs) at high rates of accretion (e.g. Ueda et
al. 2009; Uttley \& Klein-Wolt in prep.). Such `extreme' high state
BHBs probably dominate the ULX population (Walton et al. 2011) yet a
significant number of ULXs can still be found at higher
luminosities. Those above 10$^{41}$ erg s$^{-1}$ are dubbed
hyperluminous X-ray sources (HLXs: Gao et al. 2003) and provide the
best evidence (Farrell et al. 2009; Webb et al. 2012, Servillat et
al. 2011; Davis et al. 2011) for a population of intermediate mass
black holes (IMBHs: Colbert \& Mushotzky 1999). Such IMBHs (with masses above those expected from direct stellar collapse: $>$100s of M$_{\odot}$) could
potentially be formed in globular clusters (Miller \& Hamilton 2002,
but see Maccarone et al. 2007), through capturing and tidally
stripping a dwarf galaxy (King \& Dehnen 2005) or mergers in young
super star clusters (Portegies-Zwart et al. 2003, 2004).

ULXs that fall between the two categories, i.e. $L_{\rm x}$ =
$\sim$3$\times$10$^{39}$ to 1$\times$10$^{41}$ erg s$^{-1}$) remain
contentious and have been proposed as possible locations for IMBHs
accreting at low sub-Eddington rates (e.g. Miller, Fabian \& Miller
2004; Strohmayer \& Mushotzky 2009). Indeed, the brightest objects in
this class, with $L_{\rm X,peak} > 5 \times 10^{40} \rm ~erg~s^{-1}$,
have demonstrated observational properties consistent with IMBHs in
the hard state (Sutton et al. 2012). However, for the less luminous
ULXs, several problems exist with this interpretation for the {\it
  entire} population (see King 2004 for a discussion of theoretical
issues related to formation); namely the emission characteristics do
not generally match the expectation for low rates of accretion where,
in the case of a BH of mass 10$^{2-5}$ M$_{\odot}$, the emission from
the disc would still peak in the soft X-ray band and so the structure
of the flow is not expected to deviate strongly from that in BHBs at
similar rates (see Zdziarski et al. 1998; Remillard \& McClintock
2006). As a result, we would expect such sources to display a hard
spectrum up to $>$ 50 keV due to thermal Comptonisation in an electron
plasma arranged in some still-undetermined geometry. Instead, ULXs up
to 2$\times$10$^{40}$ erg s$^{-1}$ generally show spectra that cannot
easily be reconciled with sub-Eddington accretion (Stobbart et
al. 2006; Gladstone, Roberts \& Done 2009) showing a spectral break
above $\sim$3~keV (recently confirmed by {\it NuSTAR} observations of
a sample of luminous ULXs: Bachetti et al. 2013; Rana et al. 2014;
Walton et al. 2013; 2014) which, in at least one source, has been unambiguously associated with Eddington-rate accretion (Motch et al. 2014). 

Obtaining a deeper and full understanding of the nature of these
sources, requires consideration of both spectral and variability
properties {\it simultaneously}, with the latter providing a
complementary set of powerful diagnostics by which to make comparisons
to better understood sources. This has proven to be valuable,
e.g. whilst the presence of quasi-periodic oscillations (QPOs) in the
lightcurves of 5 ULXs (to date) have been used as evidence in support
of IMBHs (Strohmayer \& Mushotzky 2009; Strohmayer et al. 2007;
Strohmayer \& Mushotzky 2003; Rao, Feng \& Kaaret 2010), the details
of the variability do not generally appear to well-match this identification
(Middleton et al. 2011; Pasham \& Strohmayer 2012 but see also Pasham, Strohmayer \& Mushotsky 2014). Notably, the
recent study of Sutton et al. (2013) has reinforced the idea of using
variability properties (the fractional variability: Edelson et
al. 2002) together with spectra to broadly characterise the properties
of ULXs. This has demonstrated an apparent dependence of variability
on spectral shape, distinctly unlike that expected from IMBH
accretion. Instead it has been argued that both the spectral
(Gladstone et al. 2009, Feng \& Kaaret 2009) and variability
properties (Heil, Vaughan \& Roberts 2009) of these contentious ULXs
can be fully explained by a model for `super-critical' accretion onto
stellar mass BHs (or equally neutron stars: King 2009; Bachetti et al. 2014), where inclination and mass accretion rate are likely
to be the key determining factors in appearance (Middleton et al.
2011, 2014; Sutton et al. 2013). In most super-critical accretion
models, a large scale-height, optically thick equatorial wind is
predicted (and reproduced in MHD simulations - Ohsuga 2007; Ohsuga \&
Mineshige 2011) to be radiatively driven from the disc from within the
`spherization radius', $R_{\rm sph}$ which can be at large radii
(depending on the mass transfer rate from the secondary: Shakura \&
Sunyaev 1973; King 2004; Poutanen et al. 2007 - P07 hereafter). Unlike
models for sub-Eddington accretion onto IMBHs, this model can
reproduce the observed spectrum with the soft emission being
associated with the wind (King \& Pounds 2003; P07) and the hard
emission originating in the innermost regions which have been stripped
of material, revealing the hot disc underneath (potentially further
distorted by turbulence, advection, spin and self-heating: Beloborodov
1998; Suleimanov et al. 2002; Kawaguchi 2003). As the wind is expected
to be optically thick ($\tau \ge$ 1: P07, at least near to the disc
plane), the scattering probability is large such that, depending on
inclination angle, we may expect geometrical beaming to amplify the
hard emission (King 2009; P07) or scatter emission out of the
line-of-sight. Should the wind be inhomogeneous/clumpy (Takeuchi,
Ohsuga \& Mineshige 2013; 2014) this scattering can theoretically
imprint variability by extrinsic means (Middleton et al. 2011) and
produce the large fractional rms seen in many sources (Heil et
al. 2009; Sutton et al. 2013). However, an explanation for how this
variability mechanism operates and depends on key system parameters,
such as inclination and mass accretion rate, has been lacking.



In this paper we build on previous theory and recent spectral-timing
analyses of ULXs (notably Sutton et al. 2013) and present a simple
model for the variability originating in a radially propagating,
inhomogeneous wind. This allows us to make a series of key predictions
for the evolution with mass accretion rate (sections 2 \& 3) which we
compare to observations (sections 4, 5 \& 6).

\begin{figure*}
\begin{center}
\begin{tabular}{l}
 \epsfxsize=8cm \epsfbox{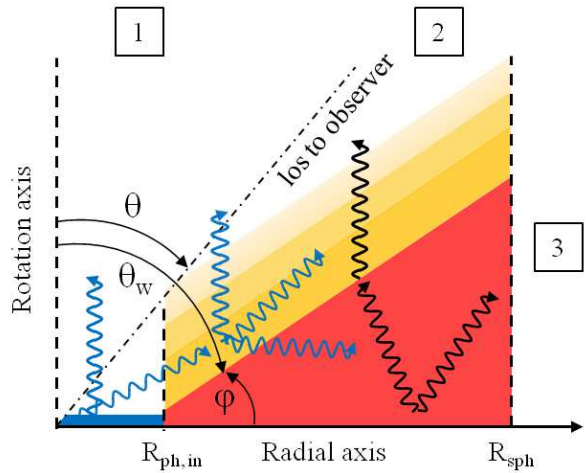}
\end{tabular}
\end{center}
\caption{Schematic representation of the super-critical model for bright ($>$ 3$\times$10$^{39}$ erg s$^{-1}$) ULXs
  (P07) where an optically thick wind is launched from the surface
  of a large scale height flow. This flow is driven from the disc from
  $R_{\rm sph}$ ($\approx$ $\dot{m}_{\rm 0}R_{\rm in}$ : Shakura \& Sunyaev 1973) down to $R_{\rm ph,in}$ within which the outflow is effectively transparent and the underlying disc emission can emerge (which may be further distorted by
  advection, spin and overheating: P07, Beloborodov 1998; Suleimanov
  et al. 2002; Kawaguchi 2003). Variability due to obscuration/scattering can be introduced when the wind is inhomogeneous - expected on large scales via
  radial propagation of flux (Lyubarskii 1997) and seen on small scales in 2D MHD simulations
  (Takeuchi et al. 2013; 2014). Depending on the orientation of the observer
  (given by $\theta$) the emission can be stochastically boosted by
  scattering into the line-of-sight (position 1) or reduced by
  scattering out of the line-of-sight (position 2). The spectrum will
  also depend heavily on $\theta$, with the emission from the hot
  inner region becoming progressively geometrically beamed with
  smaller $\theta$ and Compton down-scattered at larger $\theta$.}
\label{fig:l}
\end{figure*}

\section{The Super-critical model of accretion}

For the benefit of the reader we now summarise the key properties of
the super-critical model for accretion, as discussed and developed by
several key authors. Models describing the super-critical inflow
(Shakura \& Sunyaev 1973; King et al. 2001; King 2004; 2009; P07;
Dotan \& Shaviv 2011) differ in their precise details but the overall
picture is one where the high mass transfer rate ($\dot{m}_{\rm 0}$ -
in dimensionless units of Eddington accretion rate) from a close
binary system results in the Eddington limit being reached at large
radii. In the case where the mass transfer rate is much greater than
the `critical' value of (9/4)$\times$Eddington (P07), an optically
thick outflow (with the escape velocity, $v_{\rm esc} > \sqrt{2GM/R}$, where $M$ is the
black hole mass and $R$ is the radial distance) can be launched from
within $R_{\rm sph}$ $\approx$ $\dot{m}_{\rm 0}R_{\rm in}$ (where
$R_{\rm in}$ is the position of the inner edge in $R_{\rm g} =
GM/c^{2}$) where the scale height of the disc (proportional to
$\dot{m}_{\rm 0}$: Shakura \& Sunyaev 1973) exceeds unity. Mass is
then lost through this outflow with the mass accretion rate decaying
approximately linearly down to $R_{\rm in}$, where the accretion rate
is locally Eddington (in the limit of no advection: P07).

The geometry of the super-critical inflow can be broadly defined by
three zones (depending on the optical depth of the outflow), as fully
described in P07 (to which we direct the interested reader) and
illustrated in Figure 1.

\begin{itemize}

\item {In the innermost regions ($R <$ $R_{\rm ph, in}$), the wind is
    essentially transparent ($\tau \le$ 1) and so the emission will
    appear as a distorted `hot disc', peaking at a characteristic
    temperature, $T_{\rm in}$ (modified by some colour temperature
    correction, spin and overheating, see Beloborodov 1998; Suleimanov
    et al. 2002; Kawaguchi 2003). }

\item{At larger radii, the outflow - seen effectively as an extension
    to the large scale-height inflow - is expected to be optically
    thick for mass accretion rates above a few times Eddington.  As a
    result of advection in this optically thick material (as the
    photon diffusion timescale is longer than the viscous timescale),
    the radial temperature profile (as seen at the last scattering
    surface at $\tau$ = 1) is broadened from $R^{-3/4}$ to $R^{-1/2}$
    (see Abramowicz et al. 1988), and the result is a smeared
    blackbody extending from $T_{\rm ph, in}$ to $T_{\rm sph}$. As the
    material in the wind is outflowing and we have assumed this is an
    extension of the inflow, the viscously dissipated energy emerges
    at larger radii ($\sim$ twice the radius at which it is generated:
    P07). However, this does not affect the temperature profile (and
    nor in principal does having the wind disconnected from the inflow
    due to stratification). }

\item{At radii greater than $R_{\rm sph}$, the optical depth of the wind falls as
    1/r (P07) such that the underlying emission begins to emerge at
    approximately the radius at which it is generated, i.e. the
    emission is a quasi-thermal black-body with a peak temperature of
    $T_{\rm ph}$, which we shall refer to hereafter as the `outer
    photosphere' (and for high mass transfer rates will peak in the UV
    rather than X-ray band - equation 38 of P07). }

\end{itemize}

\subsection{Effect of a static wind}

The detailed physical properties of the wind require the global
geometry and radiative transfer to be fully modelled. Whilst beyond
the scope of this work, we can already make some simple
deductions. The large scale-height flow and wind will naturally
subtend a large solid angle to the hottest inner regions (Shakura \&
Sunyaev 1973, P07) and so a correspondingly large fraction of the high
energy flux will illuminate the material.  The latest 2D MHD
simulations (Takeuchi et al. 2013; 2014) show that the optically thick
wind will have inhomogeneities due to the Rayleigh-Taylor (or other
hydrodynamic instability) with a size of order 10~$R_{\rm s}$ ($R_{\rm
  s} = 2GM/c^{2}$). Given typical particle densities expected for these clumps
(up to $\sim10^{17}$ cm$^{-3}$: Middleton et al. 2014), the wind will
have a highly photo-ionised skin (log$\xi>$ 6 out to hundreds of
$R{\rm_g}$ for a 10 $M_{\odot}$ BH) down to an optical depth of unity
(where most of the scatterings occur). As a result, we expect the
surface material of the wind (clumpy or otherwise) in the inner
regions to be approximated to an electron plasma with a large optical
depth ($\tau\ge$ 1 - see P07) such that large amounts of incoming flux
from the inner regions will be scattered into the cone of the wind.
This naturally results in geometrical beaming (proportional to
$\dot{m}_{\rm 0}^2$: King \& Puchnarewicz 2002; King 2009) where, from
simple inspection of the assumed geometry (a hemispherical wind of
opening angle $\theta_{\rm w}$, see Figure 1), we expect a scattered
fraction of $S_{\rm f}\approx cos\theta_{\rm w}/(1-cos\theta_{\rm
  w}$). As an example, for a wind launching angle of 45\degree\
(Ohsuga et al. 2011) we would then expect a factor of $\sim$2.4 more
flux to be scattered to an observer viewing down the cone of the wind.

Scattering will not only amplify the emission from the innermost
regions when looking into the cone of the wind but also change the
energy of the scattered photons: as the wind plasma temperature is
less than that of the hot disc, i.e. $h\nu > kT_{\rm e}$, the highly
energetic incident photons will lose energy to the plasma in the
inflow via Compton down-scattering (Sunyaev \& Titarchuk 1980) and to
the outflowing wind via both recoil and bulk effects (e.g. Titarchuk
\& Shrader 2005; Laurent \& Titarchuk 2007).  At larger inclinations
(i.e. sight lines intercepting the wind), the effect of beaming
diminishes and the number of scatterings towards the observer - and
therefore energy loss - will increase (as the optical depth to the observer will be larger), although this will be complicated by the
density profile of the wind which will not be constant as radiation
momentum transfer (plus the lack of hydrostatic equilibrium) will
cause the initial clumps to expand and density drop as they move away
from the launching
point (e.g. Takeuchi et al. 2013). 

\begin{figure*}
\begin{center}
\begin{tabular}{l}
 \epsfxsize=8cm \epsfbox{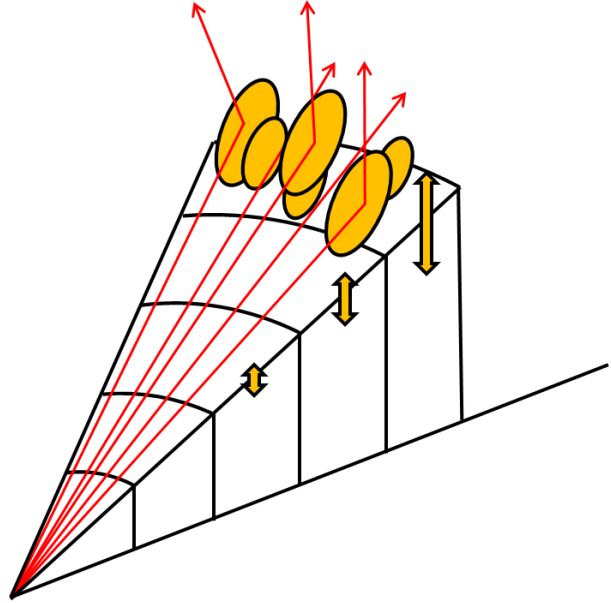}
\end{tabular}
\end{center}
\caption{Schematic slice of the inflow and wind. At large radii close
  to $R_{\rm sph}$ the wind loss is affected by propagated variability
  (Lyubarskii 1997) through the sub-critical disc into the large scale-height inflow. As the mass loss does not
  conserve the variability but instead removes it from the flow, the
  next inner radius has less with which to convolve (vertical
  arrows). Therefore within $R_{\rm sph}$ ($\approx$ $\dot{m}_{\rm 0}R_{\rm in}$), 
  the propagated variability (which can emerge from both the hot inner disc if not geometrically thin and via scattering of this radiation) is damped. Scatterings by individual clumps (Method 1) - which can imprint an extra source of variability on
  relatively short physical timescales - will average out for a large number of
  elements, i.e. at small $\theta$, but at larger $\theta$ an imprint
  is expected to remain.}
\label{fig:l}
\end{figure*}

\subsection{On the origin of variability in ULXs}

The toy model of Middleton et al. (2011) ascribed the observed large
amplitude, short timescale (tens to thousands of seconds) variability
seen in NGC~5408~X-1 to obscuration by individual clumps, generated
through radiative-hydrodynamic instabilities in the wind (now
reproduced in simulations: Takeuchi et al. 2013; 2014). In the
specific case of NGC~5408~X-1, this requires our line-of-sight to the
inner regions to intercept the wind (see Middleton et al. 2014). In
general, however, we may expect the variability in ULXs to originate
through {\it two} mechanisms and
below we provide the framework for these.

\subsubsection{Method 1: clumps}

A natural source of variability derives from the physical timescales
of the clumps (how long they exist before expanding and become
optically thin, how quickly they are launched and how large they
are). We can make a crude determination of the shortest timescales
available and the impact on the flux, by considering a single element
launched at a radial distance $R$ which remains optically thick for
the duration of the transit across the projection of the illuminating,
central region. As the distance to the observer is very large, the
crossing distance and area of the inner region is simply:
\begin{equation}
D \sim 2R_{ph, in}cos\theta\\
\end{equation}
\begin{equation}
A \sim \pi R_{ph, in}^{2}cos\theta 
\end{equation}

Assuming a maximum velocity in the perpendicular direction to the
line-of-sight equal to the escape velocity ($v_{\rm esc}$ = $\sqrt
{2GM/R}$), we find a crossing time, $t_{\rm c}$:
\begin{equation}
t_{c} \sim R_{ph, in}\sqrt{2R} \frac{GM}{c^{3}} cos\theta
\end{equation}

Where $R_{\rm ph, in}$ and $R$ are given in units of gravitational
radii ($R_{\rm g} = GM/c^{2}$).This crossing time is the fastest
timescale we should expect from this process. As an example, if we
take values of $R$ = 1000$R_{\rm g}$, $R_{\rm ph, in}\approx$
20$R_{\rm g}$ (from equation 32 of P07, this would be for a small
fraction of energy lost to the wind), $M$ = 10-100~M$_{\odot}$,
$\theta$ = 45\degree\, we find the fastest timescales to be of the
order of seconds or faster. This is of course assuming only a single
clump which, whilst not an accurate representation, remains an
illustrative limit.

It is then useful to see that the maximum drop in flux ({\it F}) we
could expect from a single, optically thick clump is simply the ratio
of the covering areas:
\begin{equation}
dF= \frac{\pi R_{c}^{2}}{A}\\
    = \left(\frac{R_{c}}{R_{ph, in}}\right)^{2}(cos\theta)^{-1}
    \end{equation}

    \noindent where $R_{\rm c}$ is the radius of the clump
    ($\sim$5$R_{\rm g}$ from simulations: Takeuchi et al. 2013). For
    the same values as above we can see that $\sim$1\% covering
    fraction is quite reasonable assuming only a single clump; in the
    limit of a greater number, this value will of course increase
    (dependent on the timescales between clumps being launched).

    In a more physical sense, N clumps will be launched at a given
    radius over a given time leading to a shot noise process and
    probability distribution. As the crossing time will more
    accurately depend on the velocity distribution (we only assumed
    the escape velocity to indicate approximate timescales) and
    details of the instabilities leading to their production (Takeuchi
    et al. 2014), we approximate the power imprinted by these events
    as a sum of zero-centred Lorentzians, cutting off at the mean of
    the velocity distribution ($v_{\rm r}$):
\begin{equation}
|\tilde{f(N_{v,r})}|^{2} \propto \frac{1}{1+[v_{r}/\bar{v_{r}}]^{2}} 
\end{equation}

\noindent where the tilde indicates a Fourier transform. We do not
define a lower frequency cutoff but we expect this to occur at the
lowest frequencies of the shot noise process (which will be a function
of the radiative hydrodynamic instabilities leading to clump
formation: Takeuchi et al. 2014). Above this low frequency break, the
sum of Lorentzians will approximate to a power-law shape with:
\begin{equation}
P(f) \propto \nu^{-\beta}
\end{equation}
\begin{equation}
\beta = \gamma(1-cos\theta), 
0 < \gamma < 2
\end{equation}

\noindent where we assume $\gamma$ to be within the range of observed
noise processes seen in accreting black hole systems (e.g. Remillard
\& McClintock 2006). Here we explicitly account for the tendency at
small inclinations for such processes to cancel out, leading only to
an increase in apparent flux to the observer (an increase as the
clumps will scatter radiation {\it towards} the observer rather than
obscure). As such, where we assume N to be large, this mechanism will
only imprint variability for ULXs at larger inclinations where the
effective number of scattering elements seen is lower due to
overlap. This is a key prediction of this model, implying that the
largest amounts of variability on the timescales of the events (which
should be relatively fast when compared to viscous changes: see next
section), should be seen for sources at higher inclinations.

Should $\dot{m}_{\rm 0}$ increase, we should expect some change in the
filling factor of the clumps. An increase may tend to lower the amount
of imprinted variability, but as the size and launching of individual
clumps (on which the variability relies) is a function of the
radiative-hydrodynamic instabilities (see Takeuchi et al. 2014), an
accurate understanding is beyond the scope of the work here.

\subsubsection{Method 2: propagating flux}

The above description assumes a wind that is `steady-state' (i.e. matter
is being launched at a constant rate at a given radius). However, in
sub-Eddington inflows, the observed (often considerable) variability
(e.g. Remillard \& McClintock 2006; Belloni 2010; Mu{\~n}oz-Darias, Motta \& Belloni
2011; Heil, Uttley \& Klein-Wolt 2014a; 2014b) originates as a result of inwardly propagating variations in
mass accretion rate/surface density through the viscous inflow
(Lyubarskii 1997; Ingram \& Done 2012) which leads to the universally
observed rms-flux relation in all accreting sources (Arevalo \& Uttley 2006; Uttley et
al. 2005, Heil \& Vaughan 2010; Heil, Vaughan \& Uttley 2012; Scaringi et al. 2012). This
can then lead to the observed PDS shape common to BHBs formed from a series
of convolved Lorentzians, damped above the local viscous timescale
(Ingram \& Done 2012).

If we assume that propagation of fluctuations still occurs in the
larger scale height super-critical flow, then this underlying and
inherent stochasticity will drive radial variations in the mass loss
in the form of the clumpy wind (see eqn 26 of P07; Figure 2). Should
the variability originate via
obscuration/scattering by clumps in the launched material, the
propagating fluctuations will imprint additional variability onto the
emission by changing the radially dependent `global' amount of mass
loss and scattering. Additionally, should the inner disc not be geometrically thin then the variability propagated down should also emerge directly. Such an origin would seem to argue against an IMBH interpretation where such discs should be geometrically thin and intrinsically stable (Churazov et al. 2001).

A clear prediction of this second mechanism is that, as mass is lost
in the wind, some fraction of the variability carried in the flow at a
given radius is expended such that the next inward radius has less
with which to convolve (Figure 2, Churazov, Gilfanov \& Revnivtsev
2001; Ingram \& van der Klis 2013) and the wind tends to radial
homogeneity with decreasing radius. The effect of this mechanism then is to
imprint the variability of the propagating flux (via scatterings or directly from the inner disc), suppressed at
frequencies higher than the local viscous timescale at $R_{\rm sph}$
(where mass loss begins):
\begin{equation}
|\tilde{\dot{M}} (r, \nu)|^{2} \propto \frac{1}{1+[\nu/\nu_{visc}(r)]^{2}}, R > R_{sph}
\end{equation}
\begin{equation}
|\tilde{\dot{M}} (r, \nu)|^{2} \propto \frac{1}{1+[\nu/\nu_{visc}(r)]^{2}} * \frac{\dot{m}}{\dot{m} + \dot{m}_w}, R \leq R_{sph}
\end{equation}

\noindent where $\dot{M}(r, \nu)$ in the above equations refers to the
mass accretion rate propagations (see Ingram \& Done 2012),
$\dot{m}_{\rm w}$ is the mass rate lost through the wind and $\dot{m}$
is the remaining accretion rate passing through the same radius (which
are also functions of $R$ and $\dot{m}_{\rm 0}$: P07).

To determine the likely timescales imprinted by this process we can
consider the viscous timescale at $R_{\rm sph}$ for representative
values, e.g. H/R (disc scale height) $\approx$ 1, $\alpha$ (viscosity parameter) = 0.01,
$\dot{m}_{\rm 0}$ = 1000-10000 (e.g. as expected for an SS433 type
system: Fuchs et al. 2003) and for a 10-100~M$_{\odot}$ black hole,
$\nu_{\rm sph}\approx$ 0.1-0.001~mHz. This is at the edge or beyond
the typically observable bandpass (due to current X-ray count rate
limitations) and so we should not expect large amounts of `rapid'
variability (e.g. tens to hundreds of seconds) to be imprinted by this
process as a result of dampening above $\nu_{\rm sph}$. However, it is
clear that even where $R_{\rm sph}$ is large, the timescales of both
methods could, in principle, overlap.

Importantly, as opposed to the contribution from single wind elements
(Method 1), at small inclinations the imprinted non-Gaussian
variability via propagated fluctuations will {\it not} average
out. The total variability for a single ULX will then be a combination
of the two methods - although at small inclinations we should only
expect variability by this second process - and a power density
spectrum (PDS) resulting from the combination. Although it may be
tempting to produce test PDS from the combination of the two methods
(specifically for those ULXs at moderate inclinations), we caution
that the relative normalisations are unknown and doing so would lead
to misleading results. In spite of the unknowns, we can make
predictions resulting from variations in $\dot{m}_{\rm 0}$ within a
combined model of the spectrum and variability as presented in the
following subsection and section 3.
\medskip

[For clarity, we reiterate that, as opposed to the origin of variability via Method 1, variability via propagating flux variations through the supercritical inflow can imprint onto the emission via the hot disc directly {\it only} if the scale height is not small (i.e. not a thin disc) and/or via scattering events from the global changes in radial mass loss in the wind (which does not rely upon the inner disc being intrinsically variable).]

\subsubsection{Effect of increasing $\dot{m}_{\rm 0}$ on Method 2}

As $R_{\rm sph}\approx$ $\dot{m}_{\rm 0}R_{\rm in}$, increasing the
mass transfer rate has a predictable effect on the observed
variability and mean energy spectrum for those sources where we view into the
evacuated funnel of the wind. We assume that the majority of the
beamed emission is from the innermost regions such that any changes in
spectral hardness (denoted in the following formulae as $h$) are a
result of the changing amount of beaming as the cone closes ($\propto$
$\dot{m}_{\rm 0}^{2}$: King \& Puchnarewicz 2002; King 2009). We also
assume that the power (fractional rms squared, from Method
2), referred to hereafter as $P$, which we approximate by a power-law
with index, $\gamma >$ 0 (i.e. not white), is seen in a fixed
frequency bandpass. As a result of increasing the mass accretion rate
from $\dot{m}_{\rm 1}$ to $\dot{m}_{\rm 2}$, we expect $R_{\rm sph}$
to increase, moving the PDS dilution break to lower frequencies,
reducing the power in our bandpass (by effectively moving the whole
PDS to lower frequencies). We can subsequently derive the expected
correlation between spectral hardness and power (for Method 2
only). From King (2009), we expect:
\begin{equation}
\frac{h_2}{h_1} = \left(\frac{\dot{m_2}}{\dot{m_1}}\right)^{2} 
\end{equation}

Assuming that all of the power is dominated by the lowest frequencies,
consistent with a decaying/diluted power above $\nu_{\rm sph}$:
\begin{equation}
\frac{P_2}{P_1} \approx \left(\frac{\nu_2}{\nu_1}\right)^\gamma
\end{equation}

Note in the above equation that the index would usually be -$\gamma$
but as the PDS is moving to lower frequencies with increasing
$\dot{m}_{\rm 0}$, across a fixed bandpass, the ratio of observed
powers is inverted. To help illustrate this point we have shown the
shift in PDS with increasing $R_{\rm sph}$ in Figure 3.

\begin{figure}
\begin{center}
\begin{tabular}{l}
 \epsfxsize=8cm \epsfbox{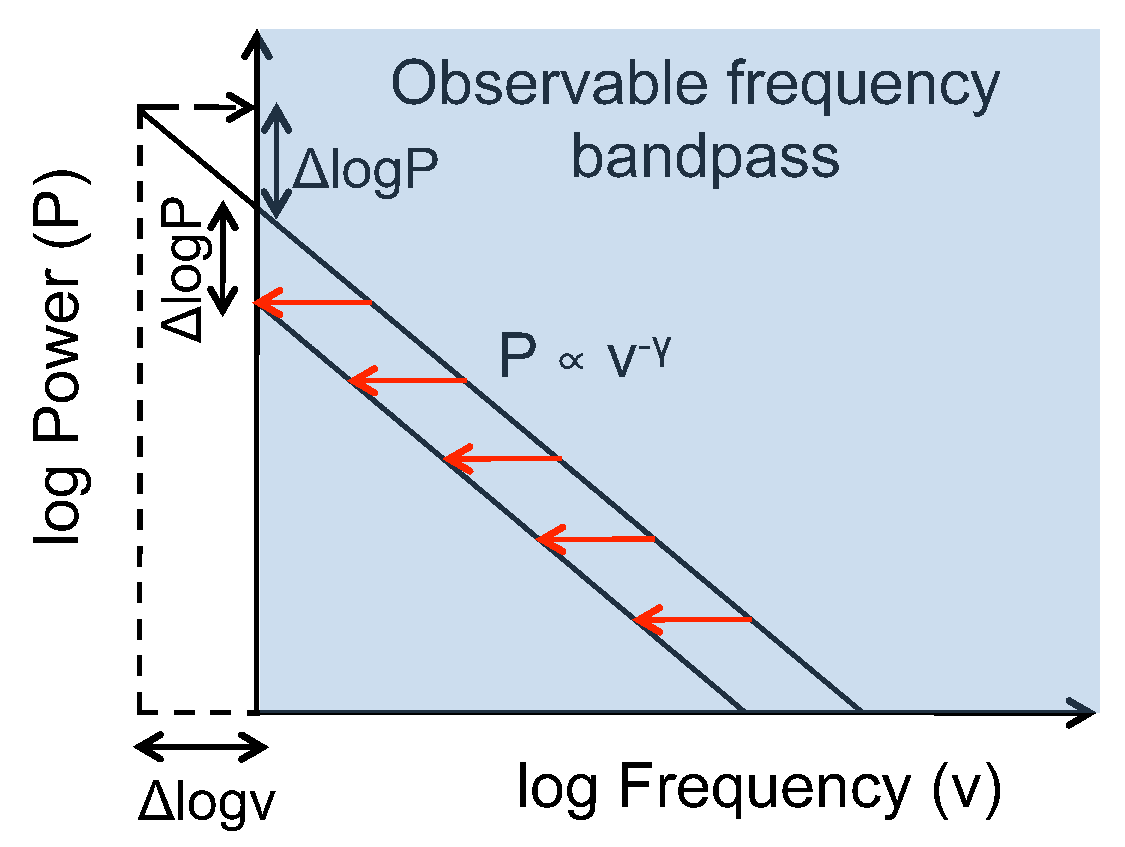}
\end{tabular}
\end{center}
\caption{Illustration of the shift of the PDS to lower frequencies
  associated with an increase in $\dot{m}_{\rm 0}$ and $R_{\rm
    sph}$. Due to the timescales at $R_{\rm sph}$, we expect to see
  the diluted power above the break (at $R_{\rm sph}$)
  where $P \propto\nu^{-\gamma}$. As the power shifts out of the
  observable bandpass (shown as the shaded area), the drop in power is
  equivalent to the increase in power that would have occurred had the
  bandpass not been fixed. As a result of the shift, we then find
  $\Delta$logP = $\gamma\Delta$log$\nu$ or $(P_{\rm 2}/P_{\rm 1}) =
  (\nu_{\rm 2}/\nu_{\rm 1})^{\gamma}$ in the formulae below, where we
  use the ratio of the viscous frequencies at each $R_{\rm sph}$ in
  the derivations for ease (and the ready connection to the change in
  $\dot{m}_{\rm 0}$).}
\label{fig:l}
\end{figure}

As we wish to consider the movement of the PDS with $\dot{m}_{\rm 0}$,
we can use the shift of the dilution break, $\nu_{\rm sph}$, which
scales as $t_{\rm dyn}^{-1}\propto$ $R^{-3/2} \propto$ $\dot{m}_{\rm
  0}^{-3/2}$:
\begin{equation}
\frac{\nu_2}{\nu_1} \approx \left(\frac{\dot{m_2}}{\dot{m_1}}\right)^{-3/2}
\end{equation} 

Combining the above, it is straightforward to see that:
\begin{equation}
\frac{P_2}{P_1} \approx  \left(\frac{h_2}{h_1}\right)^{-3\gamma/4}
\end{equation} 

By taking logarithms we then obtain: 
\begin{equation}
\Delta logP  \approx  \frac{-3\gamma}{4} \Delta logh
\end{equation} 

We note that, in the case of the above, $\gamma$ is intuitively also a
function of $\dot{m}_{\rm 0}$ as the effect of dampening is expected
to scale with $\dot{m}_{\rm w}$ (see equation 26 of P07). Whilst there
are several assumptions, it is clear that the general trend between
spectral hardness and power will be an anti-correlation when
considering only Method 2, i.e. specifically for those ULXs viewed at
smaller inclinations.

As a final caveat, we note that we have only considered the variability
to be a modulation of the high energy emission from the hot inner
disc (via scattering and/or directly), however, stochastic variability via Method 2 may also lead to
some variability of the intrinsic wind emission (the likely advection
dominated flow from $R_{\rm sph}$ to $R_{\rm ph,in}$) by varying the
column density of material and thereby changing the colour temperature
correction ($f_{\rm col}$). Although this too will tend to average out
at small $\theta$ for individual elements (Method 1), the longer
timescale trends introduced by the radial inhomogeneities will leave a
global imprint which, once again, should not average out. However,
assuming that the soft emission has the same physical origin in all
ULXs, then the lack of variability observed in NGC 5408 X-1 at these
energies (Middleton et al. 2011), implies that such variability is
likely to be relatively weak.

\section{A combined spectral-timing model}

In order to make a set of predictions for how the properties of
ULXs should evolve, we must consider {\it both} of the methods
discussed above and the spectral-variability patterns we should expect
to result from changes in $\dot{m}_{\rm 0}$. The predictions which
follow are, by necessity, only semi-quantitative; a full and accurate
quantitative picture can only be obtained from full radiative
simulations which include the nature and impact of instabilities on
the short timescale variability (and are beyond the scope of this
work).

For the following predictions, we assume that the inflow is
super-critical (or is responding in the manner of being such:
Bisnovatyi-Kogan \& Blinnikov 1977) with the wind launched from the
disc such that increasing $\dot{m}_{\rm 0}$ will still increase the
scale-height of the inflow further. We also assume that emission from
the sub-critical disc beyond the wind is negligible (i.e.
$\dot{m}_{\rm 0}$ is large enough that emission beyond $R_{\rm sph}$
is out of the X-ray bandpass).

\subsection{Source/population evolution: spectral hardness vs variability}

The inclination of the observer's line-of-sight and how this intercepts the wind ($\theta$ in Figure
1) is pivotal to the observed spectrum and power.  We
therefore dispense with past descriptive terms for ULXs based on the
spectral shape alone and instead present an inclination dependent
description based on the line-of-sight indicated in Figure 1.
\\

{\bf Sources at small $\theta$:}
\smallskip

At small inclinations, where the wind does not enter the line-of-sight (position 1 in Figure 1), the
observer sees the maximum unobscured emission from the hot inner
`disc' as well as the scattered flux from the cone of the wind. We
therefore expect a spectrum with a strong hard component with a
beaming factor scaling as $\dot{m}_{\rm 0}^2$ (King \& Puchnarewicz
2002; King 2009). Such a spectrum would correspond to the `hard
ultraluminous' class in Sutton et al. (2013).

At low inclinations we expect variability only through Method 2 with the variability emerging directly in the disc (if not geometrically thin) and/or via scattering of hot photons from the inner regions {\it into} the
line-of-sight. Given the nature of the plasma, some proportion of the 
variability should be shifted to energies below the high-energy peak via down-scattering in
the surface plasma of the wind (Titarchuk \& Shrader 2005). We may
also see variability on similar timescales in the intrinsic wind
emission possibly due to the changing radial density profile and therefore
$f_{\rm col}$, although as scattering events are predicted to dominate
the emission (via beaming), we might not expect this component to
dominate the fractional variability (although we may still detect it's presence
in an absolute variability spectrum).

At these inclinations, an increase in $\dot{m}_{\rm 0}$, which leads
to a larger $R_{\rm sph}$ and also smaller wind cone opening angle
($\theta_{\rm w}$ in Fig 1, as scale-height scales with $\dot{m}_{\rm
  0}$: Shakura \& Sunyaev 1973), can lead to one of two possibilities
(assuming no system precession):
\begin{itemize} {\item The wind remains out of the line-of-sight such
    that the spectrum will get increasingly beamed (i.e. brighter and
    harder) and the variability will drop as discussed in section
    2.2.3. The intrinsic shape of the hot component should vary little
    (P07).}  {\item The wind enters our line-of-sight such that the
    spectrum becomes softer (as hard emission is both scattered away
    and down-scattered through the wind).}
\end{itemize} 

In both cases, as $R_{\rm sph}$ increases with $\dot{m}_{\rm 0}$, we
might expect a corresponding decrease in $T_{\rm sph}$ (equation 37 of
P07), leading to a predicted anti-correlation with luminosity (see
P07, King 2009). However, contributions to the soft emission by
down-scattering and advection will complicate the evolution and this
remains an important, unresolved issue (see Miller et al. 2013, 2014).
\\

{\bf Sources at moderate $\theta$:}
\smallskip

At larger $\theta$, the wind will start to enter the line-of-sight (position 2 in Figure 1). As a result, a significant fraction of
the hard X-ray flux will be scattered out of the line-of-sight such that
the spectrum has a relatively smaller contribution from the hot inner
regions, and that which does arrive is expected to be down-scattered
such that the peak temperature of the hot component is cooler than when seen more directly, i.e. at small $\theta$. Such a spectrum would correspond to the `soft
ultraluminous' class in Sutton et al. (2013).

As opposed to ULXs at small $\theta$, the variability can originate by
both methods presented in section 2 (including possible changes in $f_{\rm }$). As the variability we can readily
observe should have a large contribution from obscuration events
(Method 1), it will likely peak at the energy of the (down-scattered) hot inner
disc. As the fraction of flux towards the observer is lower, the
integrated fractional variability (and therefore power)
should be larger than for smaller $\theta$.

We note that whilst it may appear incongruous to have both obscuration
and down-scattering, given the covering fraction of the wind to the
hot inner regions, it seems inescapable that down-scattering will
occur in some less optically thick phase of this wind (we speculate
that this could occur in the expanded clumps further from the
launching point).

\begin{figure*}
\begin{center}
\begin{tabular}{l}
 \epsfxsize=18cm \epsfbox{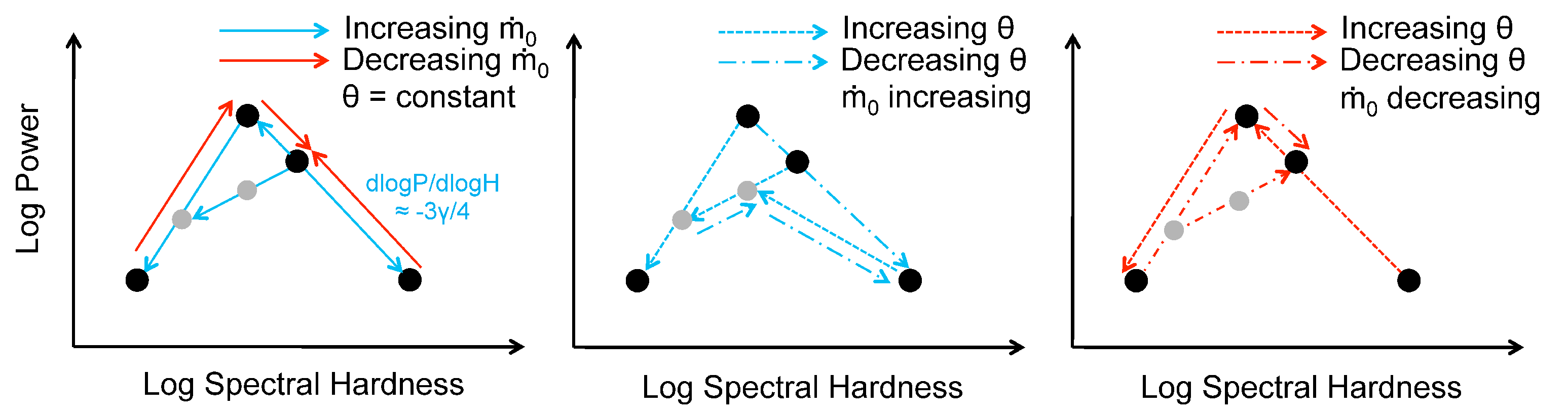}
\end{tabular}
\end{center}
\caption{Simple schematics showing the possible evolutionary paths of a ULX in spectral
  hardness (in an observing band which includes both the soft and hard
  components) and power (fractional rms squared), which depend on the
  inclination the wind makes to the observer and the change in $\dot{m}_{\rm 0}$ as described in section 3. {\it Left:} the evolution {\it without} precession is shown. When viewed at low inclinations to the wind
  (i.e. when the wind is not directly in the line-of-sight) the source is predicted to get harder with increasing $\dot{m}_{\rm 0}$, with a drop in variability due to larger $R_{\rm sph}$ and
  increased dampening in a fixed frequency bandpass (giving a gradient in log space
  of $\approx$-3$\gamma$/4 - equation 14). Alternatively, should the wind start to intercept our
  line-of-sight, the spectrum will become softer due to beaming of hard emission away from, and
  down-scattering towards, the observer, with variability
  increasing via Method 1 (where N
  elements contributing to such events is lower than at small
  inclinations). Eventually the wind will dominate, leading
  to substantial Compton down-scattering and a softer, less variable
  spectrum. The grey points indicate the evolution where an increase in mass accretion rate results in the wind tending towards homogeneity and suppressing variability via Method 1. {\it Centre:} The evolution with increasing $\dot{m}_{\rm 0}$ including the effect of precession which can further change our view of the ULX. {\it Right:} The evolution including precession with decreasing $\dot{m}_{\rm 0}$. Across a population we would expect the combination of the possible tracks, weighted by likelihood. In effect we expect this to be an inflected shape (peaking with those soft ULXs seen at moderate inclinations with large amounts of variability) with substantial scatter due to the effect of precession and wind tending towards homogeneity at larger $\dot{m}_{\rm 0}$ (suppressing variability in Method 1).  Importantly, above the softest, most variable sources we would expect a negative gradient, distinctly unlike the positive gradient expected from BHBs scaled to IMBH masses (see Belloni 2010; Mu{\~n}oz-Darias et al. 2011).}
\end{figure*}

Unlike those ULXs viewed at small inclinations, we expect only one outcome from increasing
$\dot{m}_{\rm 0}$ (once again assuming no system precession):

\begin{itemize} {\item As we expect the wind to increasingly dominate
    our sight-lines, more of the hard emission is beamed out of the
    line-of-sight whilst that in the observer's direction is
    increasingly Compton down-scattered. As a result, emission from
    the soft component should increase by a larger amount than the
    hard component. Variability via Method 2 will be increasingly
    damped, whilst the effect on Method 1 (clumps) is somewhat unclear
    (and will be a function of the generation of the instabilities)
    however, we predict that the filling factor will likely increase
    with increasing $\dot{m}_{\rm 0}$ and so we might expect a
    corresponding decrease in this contribution as the variability
    will increasingly tend to average out. Irrespective of the method,
    any variable, scattered emission from the hot disc must also pass
    through the larger optical depth inflow/outflow where reprocessing
    will redistribute it to lower energies (as with the inner disc emission) and, as we assume the soft
    emission from the wind to be relatively stable, dilute the
    fractional imprint (and therefore power). The most extreme result
    for this evolution is the spectrum and variability tending towards
    those predicted for largest $\theta$; as a corollary, it is a
    distinct possibility that the luminosity could drop below that
    considered to indicate a `bright' ULX ($>$ 3$\times$10$^{39}$ erg
    s$^{-1}$).  }
\end{itemize}

{\bf Sources at large $\theta$:}
\smallskip

In the case where the observer is viewing at largest $\theta$
(position 3 in Figure 1), the outer photosphere of the wind at $R >
R_{\rm sph}$, is predicted to dominate the observed emission (at $T_{\rm ph}$;
P07), which may be out of the X-ray bandpass and emit in the UV. Observing at such inclinations may then explain the extremely
bright UV emission seen in ultraluminous UV sources (e.g. Kaaret et
al. 2010), the super-soft ULX, M101 (Kong \& Di Stefano 2005; Shen et al. 2014), and the UV excess of SS433 (Dolan et al. 1997). As the X-ray luminosity is
expected to be low, the source may not fall into the empirical
class of bright or even faint ULXs (see as possible examples Soria et al. 2010; 2014). Whilst we may not expect these to feature in
ULX population studies, they represent an important set of `hidden'
ULXs, should provide interesting diagnostics for the winds (Middleton
\& Maccarone in prep.) and are a necessary component of any model
which hopes to describe the entire population. Although these sources are unlikely to feature in X-ray
studies (although over time we may be able to see some sources evolve to being UV bright), it is useful to make predictions about the variability properties and effect of changing $\dot{m}_{\rm 0}$. Due to the large optical depth in this direction (P07), we should not expect contributions to the variability from that emerging directly from the inner regions or via scattering and propagating through the inflow (Method 2). However, we should expect variability on the long viscous timescales at $R_{\rm sph}$. For
an observer already at large $\theta$, increasing $\dot{m}_{\rm 0}$
will increase the soft emission only.\\

{\bf Summary}
\smallskip

Here we summarise the predicted spectrum and variability properties for
a source viewed at a given inclination to the wind:

\begin{itemize}
\item{Small $\theta$: the spectrum is expected to be hard with
    variability via Method 2 only. If $\dot{m}_{\rm 0}$ increases then the
    spectrum should get harder with a corresponding drop in
    variability. Should the wind enter the line-of-sight then the
    spectrum may soften and, assuming $\dot{m}_{\rm 0}$ is not too
    high to make the wind homogeneous, the variability may increase -
    analogous to viewing at moderate $\theta$.}    
\item{Moderate $\theta$: the spectrum is expected to be soft with
    variability via Method 1 and 2 but likely dominated by Method 1 (although we re-stress that the relative normalisations are unknown at this time). If
    $\dot{m}_{\rm 0}$ increases, then the spectrum should get softer
    and the variability should decrease, both due to the wind tending
    towards homogeneity and the variability being suppressed by the stable soft component.}
\item{Large $\theta$: The spectrum should be extremely soft and
    possibly out of the X-ray bandpass altogether. Long timescale variability only should be present.}
\end{itemize}

A highly simplified version of the
predicted evolution for the population is presented in Figure 4 with
spectral hardness (in an observing band which includes both the soft
and hard components) plotted against fractional rms squared (in log
space and under the reasonable assumption that the two methods of
imprinting variability overlap in timescales: see section 2). The
overall shape is an inflection (positive to negative gradient with
spectral hardness), peaking where the sources are soft and highly
variable, corresponding to those seen at moderate inclinations. From
this inflection point, the variability drops as the spectrum hardens
(dominated by those face-on sources: see equation 14 for a derivation
of this in a single source) and drops towards softer spectral colours
(as the wind increasingly dominates the emission). The grey points
track the possible evolution of the spectrum softening out of a hard state
whilst the variability does not substantially increase (or even drops),
as a result of the wind tending towards homogeneity as $\dot{m}_{\rm
  0}$ increases. In practice this will lead to increased scatter
(depending on the likelihood of the situation) around the inflected
path. We note that the predicted shape is distinctly {\it unlike} that expected from
scaling the hardness-variability evolution of BHBs (Belloni 2010; Mu{\~n}oz-Darias, Motta
\& Belloni 2011) to IMBH masses where M$_{\rm BH}$ $>$ 100s of M$_{\odot}$ would imply sub-Eddington states and increasing rms with spectral hardness.

\subsection{Source/population evolution: effect of precession}

In the previous subsection, we made predictions for how a source's
spectral and variability properties should appear for a given
inclination and how they might change in response to increasing
$\dot{m}_{\rm 0}$ {\it without precession}. However, precession - as
seen in SS433 (see Fabrika 2004 for a review) - due to a
`slaved-disc', i.e. one where the donor star's rotational axis is
inclined to its orbital axis (Shakura 1972; Roberts 1974; van den
Heuvel et al. 1980; Whitmire and Matese 1980), must lead to changes in
the spectral and variability properties. In the simplest case, where
$\dot{m}_{\rm 0}$ remains unchanged as the disc precesses, we should
see evolution between each of the inclinations minus the effect of
increased accretion rate on homogeneity of wind and position of
$R_{\rm sph}$. Changes in $\dot{m}_{\rm 0}$ {\it as well as} precession
would lead to greater complexity but can be deduced from the
discussions above:

\begin{itemize}
\item{Low$\to$moderate $\theta$ with increasing $\dot{m}_{\rm 0}$: as
    discussed in the subsection above, {\bf if} the new $\dot{m}_{\rm
      0}$ is high enough to suppress Method 1 then the variability could
    drop whilst the spectrum softens. Otherwise, the variability will
    increase as the spectrum softens. The reverse evolution will occur
    for moderate$\to$low $\theta$ with decreasing $\dot{m}_{\rm 0}$.}
\item{Low$\to$moderate $\theta$ with decreasing $\dot{m}_{\rm 0}$:
    variability increases as the spectrum softens. The reverse
    evolution will occur for moderate$\to$low $\theta$ with increasing
    $\dot{m}_{\rm 0}$.}

\item{Moderate$\to$high $\theta$ with increasing $\dot{m}_{\rm 0}$:
    the spectrum will soften and the variability will decrease (and start to shift to longer timescales).}
\item{Moderate$\to$high $\theta$ with decreasing $\dot{m}_{\rm 0}$:
    the spectrum will soften and variability is also likely to decrease
    (or decrease with a flatter gradient).}
\end{itemize}

We plot the possibilities above for both increasing and decreasing $\dot{m}_{\rm 0}$ in Figure 4.
There are clearly several paths a given ULX may take
depending on the presence of precession, $\dot{m}_{\rm 0}$ and starting inclination,
with the overall trend we should expect to observe across the {\it
  population} resulting from the combination of the possibilities
weighted against their likelihood. Whilst precession will no doubt distort observational trends, by studying a large number of sources we can expect to average across the effect and still see an inflected shape across the population.

\begin{table*}
\begin{center}
\begin{minipage}{134mm}
\bigskip
\caption{ULX sample observational information}
\begin{tabular}{l|c|c|c|c|c|c}
  \hline

ULX & OBSID & obs. date & useful exposure & $L_{\rm x}$ & count rate \\
 &      &   & (ks)  & ($\times 10^{39} \rm ~erg~s^{-1}$)
& (ct s$^{-1}$) \\
   \hline
NGC 5408 X-1   & 0302900101  & 2006-01-13 & 85.4 & 4.8 & 1.0  \\
4.85 Mpc   &  0500750101 & 2008-01-13 & 28.6 & 4.5 &  0.9 \\
   &  0653380201 & 2010-07-17 & 71.8 & 5.6 &  1.2 \\
   &  0653380301 & 2010-07-19 & 88.2 & 5.6 &  1.1 \\
   &  0653380401 & 2011-01-26 & 73.4 & 5.3 &  1.1 \\
   &  0653380501 & 2011-01-28 & 69.2 & 5.3 &  1.0 \\  
NGC 6946 X-1   &  0200670101 & 2004-06-09 & 2.1 & 3.8 &  0.3 \\
 5.96 Mpc  & 0200670301  & 2004-06-13 & 10.0 & 3.0 & 0.3  \\
   & 0200670401  & 2004-06-25 & 3.2 & 3.8 & 0.3  \\
   & 0401360301  & 2006-06-18 & 2.7 & 3.4 &  0.3 \\
   & 0500730101  & 2007-11-08 & 17.4 & 3.0 & 0.2  \\
   & 0500730201  & 2007-11-02 & 24.5 & 2.6 &  0.2 \\
   & 0691570101  & 2012-10-21 & 81.1 & 3.8 &  0.4 \\
NGC 5204 X-1   & 0142770101  & 2003-01-06 & 13.9 & 4.8 & 0.6  \\
5.15 Mpc   & 0142770301  &  2003-04-25 & 4.1 & 6.4  & 0.8  \\   
   & 0150650301  &  2003-05-01 & 4.6 & 6.7 &  1.0 \\
   & 0405690101  &  2006-11-15 & 1.7 & 7.9 & 1.2  \\
   & 0405690201  &  2006-11-19 & 30.4 & 7.3 & 1.0  \\
   & 0405690501  &  2006-11-25 & 20.9 & 6.0 & 0.8  \\
NGC 1313 X-1/X-2   & 0106860101  & 2000-10-17 & 11.9 & 4.9/1.5 & 0.7/0.2  \\
3.95 Mpc   & 0150280301  & 2003-12-21 & 7.0 & 8.6/6.4 & 1.0/0.9 \\
   & 0150280401  & 2003-12-23 & 3.0 & 6.4/7.5 &  0.7/1.0 \\
   & 0150280501  & 2003-12-25 & 6.6 & -/2.8 &  -/0.5 \\
   & 0150280601  & 2004-01-08 & 8.2 & 7.7/2.2 &  0.8/0.4 \\
   & 0205230301  & 2004-06-05 & 8.6 & 6.7/7.5 &  1.0/1.0 \\   
   & 0205230401  & 2004-08-23 & 3.8 & 3.0/1.7 &  0.6/0.3 \\
   & 0205230501  & 2004-11-23 & 12.5 & -/1.7 &  -/0.3 \\
   & 0205230601  & 2005-02-07 & 8.4 & 5.2/7.1 &  0.6/1.0 \\
   & 0301860101  & 2006-03-06 & 16.6 & -/6.5 &  -/0.7 \\
   & 0405090101  & 2006-10-15 & 74.0 & 4.9/6.2 &  0.7/0.7 \\
Ho II X-1   & 0112520601  & 2002-04-10 & 4.6 & 8.5 & 3.2 \\
 3.34 Mpc  & 0112520701  & 2002-04-16 & 2.1 & 7.7 & 2.8  \\
   & 0112520901  & 2002-09-18 & 3.5 & 1.9 & 0.8  \\
   & 0200470101  & 2004-04-15 & 22.2 & 9.1 & 3.1  \\
   & 0561580401  & 2010-03-26 & 21.0 & 3.3 & 1.3  \\
Ho IX X-1   & 0112521001   & 2002-04-10 & 7.0 & 14.6 & 1.9  \\   
4.23 Mpc   & 0112521101  & 2002-04-16 & 7.6  & 16.7 & 2.2 \\
   & 0200980101  & 2004-09-26 & 57.2 & 12.6 & 1.5  \\
   & 0657801801  & 2011-09-26 & 6.9  & 23.1 & 2.5  \\
   & 0657802001  & 2011-03-24 & 2.7  & 16.7 & 1.4  \\
   & 0657802201  & 2011-11-23 & 12.6 & 22.1 & 2.3  \\
IC 342 X-1   & 0093640901  & 2001-02-11  & 4.8 & 3.4 & 0.3  \\
3.61 Mpc   & 0206890101  & 2004-02-20 & 6.8 & 7.5 & 0.9  \\
   & 0206890201  & 2004-08-17 & 17.1 & 3.9 & 0.4  \\
   & 0206890401  & 2005-02-10 & 3.3 & 11.0 &  1.2 \\   
   NGC 55 ULX-1 & 0028740201 & 2001-11-14 & 30.4 &1.5 & 1.4\\
   1.95 Mpc & 0655050101 & 2010-05-24 & 56.6 & 0.6 & 0.8\\
   \hline

\end{tabular}
Notes: Observational properties of the sample of ULXs studied in this
paper including the {\it XMM-Newton} observation ID, date of exposure,
duration of spectral exposure (after accounting for background flares
and deadtime), 0.3-10~keV X-ray luminosity ({\it absorbed}, determined from the model as per the text) using distances from NASA NED (http://ned.ipac.caltech.edu) provided below the source name, and EPIC-PN count
rate. In the case of NGC 1313, ``-" indicates that X-1 was outside of the detector's field-of-view when the observation was taken.
\end{minipage} 

\end{center}
\end{table*}

\begin{table*}
\begin{center}
\begin{minipage}{176mm}
\bigskip
\caption{Model fitting parameters and $F_{\rm var}$ values}
\begin{tabular}{|l||c|c|c|c|c|c|c|c}
  \hline

Obs.ID & $N_{\rm H}$ & $kT_{\rm d}$ & $\Gamma$ & $kT_{\rm e}$ & $f_{\rm 0.3-1}$ & $f_{\rm 1-10}$ & $\chi^2$/d.o.f & $F_{\rm var}$\\
  & ($\times$10$^{22}$ cm$^{-2}$) & (keV) & & (keV)  &  \multicolumn{2}{c}{($\times$10$^{-12}$ erg s$^{-1}$ cm$^{-2}$)} & &  (\%)\\
   \hline\\
\multicolumn{9}{c}{NGC 5408 X-1}\smallskip\\

0302900101  & 0.10 $\pm$ 0.01  & 0.18 $\pm$ 0.01 & 2.35 $\pm 0.12$ & 1.70 $^{+0.55}_{-0.29}$ &  1.85 $\pm$ 0.04  & 0.85 $\pm$ 0.02 & 670/525 & 21 $\pm$ 1\\

   0500750101 &  0.11 $^{+0.02}_{-0.01}$  & 0.18 $\pm$ 0.01 & 2.23 $\pm$ 0.19 & 1.57 $^{+1.01}_{-0.36}$ & 1.71 $\pm$ 0.07 & 0.87 $\pm$ 0.03 & 398/356 & 22 $\pm$ 1\\ 

   0653380201 & 0.09 $\pm$ 0.01 & 0.20 $\pm$ 0.01 & 2.21 $\pm$ 0.11 & 1.55 $^{+0.33}_{-0.21}$ & 1.84 $\pm$ 0.04 & 1.12 $\pm$ 0.02 & 686/552& 19 $\pm$ 1\\ 

   0653380301 & 0.09 $\pm$ 0.01  & 0.20 $\pm$ 0.01 & 2.27 $\pm$ 0.10 & 1.80 $^{+0.50}_{-0.28}$ & 1.71 $\pm$ 0.04 & 1.08 $\pm$ 0.02 & 682/581& 19 $\pm$ 1\\ 

   0653380401 & 0.10 $\pm$ 0.01  & 0.19 $\pm$ 0.01 & 2.28 $\pm$ 0.12 & 1.69 $^{+0.54}_{-0.29}$ & 1.78 $^{+0.04}_{-0.05}$ & 1.00 $\pm$ 0.02 & 575/535& 20 $\pm$ 1\\ 

   0653380501 & 0.09 $\pm$ 0.01 & 0.20 $\pm$ 0.01 & 2.25 $^{+0.13}_{-0.12}$ & 2.16 $^{+1.64}_{-0.50}$ & 1.58 $\pm$ 0.04 & 1.03 $\pm$ 0.02 & 534/531 & 20 $\pm$ 1\\ 
\\
\multicolumn{9}{c}{NGC 6946 X-1}\smallskip\\

  0200670101 & $<$0.27 & 0.34 $^{+0.05}_{-0.12}$ & $<$ 1.99 & $>$ 0.88 & 0.63 $^{+0.53}_{-0.08}$ & 0.79 $^{+1.13}_{-0.18}$ & 15/19& 22 $\pm$ 7\\ 

    0200670301  &  0.25 $^{+0.03}_{-0.05}$  & 0.30 $^{+0.06}_{-0.04}$ & $<$ 1.85 & $<$ 2.18 & 0.63 $\pm$ 0.03 & 0.62 $\pm$ 0.03 & 81/93& (23 $\pm$ 8)\\ 
   
    0200670401  &   0.33 $^{+0.14}_{-0.09}$ & 0.23 $^{+0.10}_{-0.08}$ & 2.23 $^{+0.46}_{-0.42}$ & unconstrained & 1.05 $\pm$ 0.09 & 0.76 $^{+0.07}_{-0.06}$ & 45/33& 33 $\pm$ 9\\ 
   
    0401360301  &  0.28 $^{+0.05}_{-0.09}$ & 0.26 $^{+0.07}_{-0.10}$ & $<$ 2.62  & unconstrained & 0.88$\pm$ 0.08 & 0.66 $\pm$ 0.06 & 18/26& (27 $\pm$ 15)\\ 

    0500730101  & 0.27 $^{+0.07}_{-0.05}$ & 0.24 $^{+0.03}_{-0.05}$ & 1.84 $^{+0.40}_{-0.21}$ & $>$ 1.14 & 0.69 $^{+0.09}_{-0.03}$ & 0.62 $^{+0.05}_{-0.02}$ & 130/128& 42 $\pm$ 4\\ 
   
    0500730201  & 0.37 $^{+0.08}_{-0.06}$ & 0.20 $^{+0.02}_{-0.03}$ & 2.04 $^{+0.36}_{-0.17}$ & $>$ 1.23 & 0.99 $^{+0.09}_{-0.06}$ & 0.58 $\pm$ 0.03 & 144/136 &  30 $\pm$ 5\\ 

     0691570101 & 0.30 $^{+0.03}_{-0.02}$  & 0.23 $\pm$ 0.02 & 1.98 $^{+0.13}_{-0.07}$ & 1.68 $^{+0.44}_{-0.26}$ &  0.99 $^{+0.04}_{-0.02}$ & 0.80 $^{+0.02}_{-0.01}$ &  598/501 & 31 $\pm$ 2\\ 

\\
\multicolumn{9}{c}{NGC 5204 X-1}\smallskip\\

 0142770101  & $<$0.05 & 0.24 $\pm$ 0.03 & 1.74 $^{+0.15}_{-0.17}$ & 1.65 $^{+0.53}_{-0.29}$ & 0.58 $^{+0.02}_{-0.01}$ & 1.11 $\pm$ 0.05 & 197/241 & $<$ 11 \\ 

    0142770301$^{**}$  &  0.05 $\pm$ 0.01 & 0.32 $^{+0.03}_{-0.04}$ & 1.70 $^{+0.15}_{-0.14}$ & $>$ 1.24 & 0.85 $^{+0.06}_{-0.05}$ & 1.33 $^{+0.13}_{-0.11}$ & 114/111& $<$ 10\\  

    0150650301  & $<$ 0.07 & 0.39 $^{+0.06}_{-0.11}$ & $<$ 2.28 & 1.42 $^{+4.21}_{-0.36}$ & 0.88 $^{+0.15}_{-0.05}$ & 1.47 $\pm$ 0.09  & 135/135 & $<$ 14\\ 

    0405690101$^{*}$  &  $<$ 0.09 &  0.31 $\pm$ 0.04 &  $<$ 2.42 & 0.87 $^{+1.71}_{-0.19}$ & 1.40 $^{+0.13}_{-0.12}$ & 1.53 $\pm$ 0.12 & 63/66 & $<$ 17 \\ 

    0405690201  &  0.06 $\pm$ 0.01 & 0.32 $^{+0.03}_{-0.04}$ & 2.11 $^{+0.30}_{-0.26}$ & $>$ 1.67 & 1.04 $^{+0.07}_{-0.06}$ & 1.57 $\pm$ 0.04 & 493/454& $<$ 6\\ 

    0405690501  &  $<$ 0.06 & 0.29 $^{+0.01}_{-0.04}$  & 1.89 $^{+0.12}_{-0.15}$ & $>$ 2.13 & 0.73 $^{+0.03}_{-0.02}$ & 1.36 $\pm$ 0.05 & 331/355 & $<$ 10\\ 

\\
\multicolumn{9}{c}{NGC 1313 X-1}\smallskip\\

 0106860101  & 0.22 $^{+0.05}_{-0.04}$ & 0.29 $^{+0.08}_{-0.07}$ & 1.64 $^{+0.13}_{-0.18}$ & 2.20 $^{+1.31}_{-0.49}$ & 0.95 $^{+0.05}_{-0.06}$ & 2.39 $\pm$ 0.09 & 265/259 & 20 $\pm$ 4\\ 

    0150280301$^{*}$  & 0.28 $^{+0.03}_{-0.02}$ & 0.41 $^{+0.05}_{-0.04}$ & 1.85 $^{+0.56}_{-0.41}$ & $>$ 1.14  & 2.02 $\pm$ 0.11 & 4.36 $^{+0.18}_{-0.17}$ & 215/209& $<$ 13\\ 

    0150280401$^{*}$  & 0.34 $\pm$ 0.05 & 0.39 $^{+0.08}_{-0.06}$ & 1.64 $^{+0.56}_{-0.61}$ & $>$ 1.02 & 1.58 $^{+0.16}_{-0.15}$ & 3.30 $^{+0.29}_{-0.25}$ & 62/69 & $<$ 20\\ 

    0150280601$^{*}$  & 0.23 $\pm$ 0.02 & 0.48 $\pm$ 0.04 & 1.66 $^{+0.25}_{-0.30}$ & $>$ 1.59 & 1.43 $\pm$ 0.06 & 3.82 $^{+0.18}_{-0.17}$ & 200/206& $<$ 10\\ 

    0205230301$^{*}$  & 0.26 $\pm$ 0.02 & 0.43 $^{+0.05}_{-0.04}$ & 1.64 $^{+0.42}_{-0.38}$ & 1.32 $^{+0.69}_{-0.25}$ & 1.42 $\pm$ 0.07 & 3.47 $\pm$ 0.12 & 239/251& $<$ 12\\  

    0205230401  & 0.30 $^{+0.10}_{-0.07}$ & 0.29 $^{+0.07}_{-0.12}$ & $<$ 2.84 & 0.65 $^{+1.13}_{-0.09}$ & 1.77 $^{+1.34}_{-0.16}$ & 1.34 $^{+0.11}_{-0.09}$ & 97/73 & (16 $\pm$ 9)\\ 

    0205230601  & 0.37 $^{+0.10}_{-0.08}$ & 0.20 $^{+0.06}_{-0.04}$ & 1.74 $^{+0.14}_{-0.17}$ & 1.93 $^{+2.09}_{-0.48}$ & 2.27 $\pm$ 0.17 & 2.69 $^{+0.16}_{-0.15}$ & 138/154 & (12 $\pm$ 8)\\ 

    0405090101  & 0.29 $\pm 0.02$ & 0.24 $\pm$ 0.02 & 1.68 $\pm$ 0.04 & 2.25 $^{+0.29}_{-0.21}$ & 1.31 $\pm$ 0.03 & 2.46 $\pm$ 0.04 & 865/793 & 17 $\pm$ 2\\ 

   \hline

\end{tabular}
Notes: Best-fitting parameters from fitting the continuum model ({\sc
  tbabs*(diskbb+nthcomp)}) with errors quoted at the 90\% level. The
column density ($N_{\rm H}$) has assumed the
abundances of Wilms et al. (2000). $kT_{\rm d}$ is the peak disc
temperature in the {\sc diskbb} component, $kT_{\rm e}$ is the
electron plasma temperature and $\Gamma$ is the photon-index (in the
{\sc nthcomp} component) connecting the seed photon temperature to the
high energy rollover at $\sim$2-3~$kT_{\rm e}$. The flux values (in
units of $\times 10^{-12} \rm ~erg~cm^{-2}~s^{-1}$) are quoted for the
unabsorbed model integrated over 0.3-1~keV and 1-10~keV respectively
by including the {\sc cflux} component. In order to obtain parameters
which can loosely be related to our model and are generally more
physically motivated, we ensure that the high energy rollover is in
the bandpass and the input seed photon temperature is fixed to the
peak of the soft component. This is then consistent with our physical
model where Compton down-scattering is likely to broaden the shape
below the peak of the hard emission. In many cases the continuum was
too broad to provide constraining fits in the required parameter space
when all parameters were free; in these cases we determined the error
bounds on each component in turn. Where the observation is highlighted
by an asterisk (by the Obs.ID), the temperature ($kT_{\rm d}$ or
$kT_{\rm e}$) and/or photon index of the component was frozen; where a
double-asterisk is given, the normalisation of the {\sc diskbb}
component was also frozen to determine errors on the {\sc nthcomp}
component. Fractional variability was determined by integrating the
PDS (from 3 to 200 mHz), with values below 3 sigma significance given
in parentheses. Stringent upper limits were determined by simulating
using the PDS of NGC~6946~X-1 as input (see section 4.2 for details).

\end{minipage} 

\end{center}
\end{table*}

\begin{table*}
\begin{center}
\begin{minipage}{176mm}
\bigskip
\caption{Model best-fitting parameters and $F_{\rm var}$ values}
\begin{tabular}{l|l|c|c|c|c|c|c|c|c}
  \hline

Obs.ID & $N_{\rm H}$ & $kT_{\rm d}$ & $\Gamma$ & $kT_{\rm e}$ & $f_{\rm 0.3-1}$ & $f_{\rm 1-10}$ & $\chi^2$/d.o.f. & $F_{\rm var}$\\
  & ($\times$10$^{22}$ cm$^{-2}$) & (keV) & & (keV)  & \multicolumn{2}{c}{($\times$10$^{-12}$ erg s$^{-1}$ cm$^{-2}$)} & & (\%)\\
   \hline\\
   \multicolumn{9}{c}{NGC 1313 X-2}\smallskip\\
 0106860101  & 0.24 $^{+0.08}_{-0.06}$ & 0.31 $^{+0.13}_{-0.10}$  & $<$ 2.21  & $>$ 1.03 & 0.47 $^{+0.10}_{-0.11}$ & 0.75 $^{+0.08}_{-0.06}$ & 97/97 & (19 $\pm$ 10) \\ 

    0150280301$^{*}$  & 0.20 $\pm$ 0.03 & 0.60 $^{+0.10}_{-0.08}$ & 1.44 $^{+0.40}_{-0.35}$ & 1.62 $^{+0.92}_{-0.29}$  & 0.62 $\pm$ 0.04 & 3.28 $\pm$ 0.12 & 173/208 & (14 $\pm$ 5) \\ 

    0150280401$^{**}$  & 0.19 $\pm$ 0.03 & 0.68 $^{+0.17}_{-0.13}$ & 1.38 $^{+0.11}_{-0.10}$  & 1.62 $^{+0.39}_{-0.23}$ & 0.67 $\pm$ 0.06 & 3.77 $^{+0.21}_{-0.20}$ & 91/103 & $<$ 16 \\ 

    0150280501  & 0.27 $^{+0.12}_{-0.07}$ & 0.30 $^{+0.21}_{-0.15}$ & $<$ 2.04  & 1.37 $^{+0.94}_{-0.40}$  & 0.71 $^{+0.65}_{-0.20}$ & 1.45 $^{+0.11}_{-0.10}$ & 131/125 & (12 $\pm$ 10) \\ 

    0150280601$^{**}$  & 0.21 $\pm$ 0.02 & 0.44 $^{+0.08}_{-0.07}$ & 1.79 $^{+0.19}_{-0.17}$  & $>$1.27 & 0.47 $^{+0.04}_{-0.03}$ & 1.11 $^{+0.08}_{-0.07}$  & 112/126 & 22 $\pm$ 6 \\ 
   
    0205230301$^{**}$  & 0.25 $\pm$ 0.02 & 0.56 $^{+0.09}_{-0.08}$ & 1.45 $\pm$ 0.06  & 1.54 $^{+0.15}_{-0.12}$  & 0.81 $\pm$ 0.04 & 3.91 $\pm$ 0.12 & 215/262& $<$ 13 \\ 
   
    0205230401$^{*}$  & 0.18 $\pm$ 0.05 & 0.38 $^{+0.10}_{-0.08}$   & $<$ 1.83 & $>$ 0.96 & 0.37 $^{+0.05}_{-0.04}$ & 0.78 $^{+0.16}_{-0.10}$ & 30/38 & (30 $\pm$ 11) \\ 

    0205230501$^{*}$  & 0.25 $\pm$ 0.03 & 0.29 $^{+0.09}_{-0.01}$  & 1.96 $^{+0.28}_{-0.24}$ & $>$ 1.40  & 0.53 $\pm$ 0.04 & 0.82 $\pm$ 0.05 & 113/142 & 28 $\pm$ 5 \\ 

    0205230601$^{*}$  & 0.19 $\pm$ 0.03 & 0.68 $^{+0.10}_{-0.08}$ & $<$ 1.83 & 1.52 $^{+0.63}_{-0.23}$ & 0.59 $\pm$ 0.03 & 3.67 $\pm$ 0.12 & 239/259 & $<$ 13 \\ 

    0301860101$^{*}$  & 0.23 $^{+0.02}_{-0.03}$  & 0.69 $^{+0.06}_{-0.05}$ & $<$ 1.61 & 1.35 $^{+0.37}_{-0.10}$ & 0.65 $\pm$ 0.03 & 3.42 $\pm$ 0.09 & 331/355 & (7 $\pm$ 7) \\ 

    0405090101$^{*}$  & 0.21 $\pm$ 0.01 & 0.67 $\pm$ 0.03 & 1.51 $^{+0.26}_{-0.22}$ & 1.68 $^{+0.42}_{-0.21}$ & 0.60 $\pm$ 0.01 & 3.14 $\pm$ 0.04 & 807/856 & $<$ 8\\ 

\\
\multicolumn{9}{c}{Ho~II X-1}\smallskip\\
 0112520601$^{*}$  & 0.07 $\pm$ 0.01 & 0.32 $\pm$ 0.02 & 2.06 $^{+0.30}_{-0.22}$ &  $>$ 1.45 &  3.29 $\pm$ 0.10 & 4.39 $\pm$ 0.15 & 325/325& 9 $\pm$ 3\\ 

   0112520701$^{*}$  &   0.06 $\pm$ 0.01 & 0.31 $^{+0.04}_{-0.03}$  & 2.06 $^{+0.31}_{-0.25}$ & $>$ 1.43 & 2.74 $^{+0.14}_{-0.13}$ & 4.08 $^{+0.21}_{-0.20}$ & 211/170& $<$ 12\\ 

    0112520901  &  0.10 $^{+0.06}_{-0.04}$  & 0.18 $^{+0.06}_{-0.05}$ & $<$ 2.90 & $>$ 0.61 & 1.33 $^{+0.13}_{-0.12}$ & 0.74 $\pm$ 0.06 & 101/86& $<$ 12\\ 
   
    0200470101$^{*}$  & 0.07 $\pm$ 0.01  & 0.32 $\pm$ 0.01 & 2.04 $^{+0.11}_{-0.10}$ & 1.77 $^{+0.42}_{-0.24}$ & 3.37 $\pm$ 0.05 & 4.73 $^{+0.06}_{-0.07}$ & 594/598& $<$ 5\\ 

    0561580401  & 0.08 $\pm$ 0.01  & 0.23 $\pm$ 0.02 & 1.91 $^{+0.20}_{-0.24}$  & 1.19 $^{+0.25}_{-0.17}$ & 1.77 $^{+0.09}_{-0.12}$ & 1.49 $\pm$ 0.04 & 411/387 & $<$ 6\\ 

\\
\multicolumn{9}{c}{Ho~IX X-1\smallskip}\\
 0112521001$^{*}$   &  0.14 $\pm$ 0.01  &  0.43 $^{+0.04}_{-0.03}$  &   1.54 $^{+0.09}_{-0.10}$ & 2.25 $^{+0.70}_{-0.36}$ & 1.64 $\pm$ 0.05 & 6.14 $\pm$ 0.17  & 350/388& $<$ 7\\   

    0112521101$^{*}$  & 0.14 $\pm$ 0.01  & 0.49 $\pm$ 0.03 & 1.60 $\pm$ 0.13 & $>$ 1.95  & 1.85 $\pm$ 0.05 & 7.06 $\pm$ 0.17 & 425/456& (5 $\pm$ 5)\\ 

    0200980101  &   0.16 $\pm$ 0.01  & 0.28 $\pm$ 0.02  & 1.55 $\pm$ 0.03 & 2.50 $^{+0.23}_{-0.18}$ & 1.30 $\pm$ 0.02 & 5.36 $\pm$ 0.06 & 1030/1042& $<$ 6\\ 

    0657801801$^{*}$  &   0.15 $\pm$ 0.01  & 0.61 $\pm 0.04$ & 1.66 $^{+0.27}_{-0.26}$ & $>1.77$  & 2.38 $\pm$ 0.04 & 9.97 $\pm$ 0.22 & 443/465 & $<$ 5\\ 

    0657802001$^{*}$  & 0.13 $\pm$ 0.02  & 0.54 $^{+0.07}_{-0.06}$ & 1.41 $^{+0.14}_{-0.26}$ & $>$ 1.73  & 1.58 $\pm$ 0.06 & 7.16 $\pm$ 0.35 & 115/123 & $<$ 13\\ 

    0657802201$^{*}$  &   0.17 $\pm$ 0.01  & 0.60 $\pm$ 0.04  & 1.63 $^{+0.19}_{-0.18}$  & 2.23 $^{+1.34}_{-0.42}$ & 2.11 $\pm$ 0.05 & 9.59 $\pm$ 0.18 & 589/622 & $<$ 7\\ 

\\
\multicolumn{9}{c}{IC~342 X-1}\smallskip\\

 0093640901$^{*}$  & 0.77 $\pm$ 0.13 & 0.51 $^{+0.16}_{-0.12}$ & 1.34 $^{+0.36}_{-0.23}$ & 1.82 $^{+2.56}_{-0.43}$ & 0.57 $\pm$ 0.07 & 2.44 $^{+0.18}_{-0.17}$ & 60/57& $<$ 25\\ 

    0206890101$^{**}$  & 0.83 $^{+0.05}_{-0.04}$ & 0.81 $^{+0.09}_{-0.08}$ & 1.32 $\pm$ 0.11 & 1.81 $^{+0.36}_{-0.23}$ & 1.16 $\pm$ 0.07 & 5.63 $\pm$ 0.19 & 187/207& (9 $\pm$ 9)\\ 
      
    0206890201$^{*}$  &   0.76 $\pm$ 0.06 & 0.62 $^{+0.06}_{-0.05}$ &  $<$ 1.51 & 1.92 $^{+0.83}_{-0.34}$ & 0.62 $\pm$ 0.03 & 2.84 $\pm$ 0.10 & 217/232 & 22 $\pm$ 4\\ 

    0206890401$^{**}$  & 0.79 $\pm$ 0.06  & 0.89 $^{+0.16}_{-0.14}$  & 1.43 $\pm$ 0.12 & 2.23 $^{+1.23}_{-0.44}$ & 1.27 $\pm$ 0.10 & 8.01 $^{+0.36}_{-0.35}$ & 132/136 & $<$ 14\\  

\\
\multicolumn{9}{c}{NGC 55 ULX-1}\smallskip\\

 0028740201 & 0.27 $\pm$ 0.02 & 0.28 $\pm$ 0.04 & $<$ 2.06 & 0.71 $^{+0.10}_{-0.07}$ & 3.45 $^{+0.38}_{-0.28}$ & 2.70 $^{+0.05}_{-0.04}$ & 532/477 & $<$ 7\\

 0655050101 & 0.33 $^{+0.02}_{-0.01}$ & 0.22 $\pm$ 0.02 & $<$ 2.04 & 0.54 $^{+0.06}_{-0.05}$ & 2.83 $^{+0.26}_{-0.21}$ & 1.11 $^{+0.02}_{-0.01}$ & 670/522 & 10 $\pm$ 3\\

   \hline

\end{tabular}
Notes: As for Table 2.
\end{minipage} 

\end{center}
\end{table*}

\section{Observational comparison}

The work of Sutton et al. (2013), goes some way to establishing a
trend of variability with spectral hardness which (encouragingly)
appears similar to our predictions. However, to rigourously test our
model predictions and identify weaknesses it is important to examine
the spectral and variability evolution of a high quality sample of
ULXs - excluding those which can now be confidently associated with
`normal' modes of BHB evolution (Middleton et al. 2013) - in greater
detail.

We select nine of the brightest ULXs (in flux), known to span a range
of spectral shapes (Gladstone et al. 2009), that have been
well-observed by {\it XMM-Newton} to ensure the availability of high
quality datasets for spectral and timing analysis. Importantly we exclude the
`broadened disc' class of sources from Sutton et al. (2013) which are considered to be BHBs experiencing normal mass transfer rates but with Eddington inflows in their most inner regions (Middleton et al. 2011; 2012; 2013). The
observational details of our sample are given in Table 1.

For all observations we re-process the data using {\sc sas v.12.0.1}
and up-to-date calibration files. We apply standard data filters and
flags for bad pixels and patterns (see the {\it XMM-Newton} user's
handbook\footnotemark\footnotetext{http://xmm.esac.esa.int/external/xmm\_user\_support\\/documentation/uhb/}),
and remove periods of high energy background (based on the 10-15~keV
count rate from the full field of view) associated with soft proton
flaring.

We use {\sc xselect} to extract EPIC-PN spectra and lightcurves from
source regions not smaller than 30 arcsec radius (with the exception
of the first 6 observations of NGC 6946 X-1 as the source is close to
a chip-gap) and background regions chosen to avoid the pixel read-out
direction and other sources in the field. We do not use the EPIC-MOS
in this analysis as the fractional increase in data is small, and the
instrumental responses differ at soft energies.

\begin{figure*}
\centering
\mbox{\subfigure{\includegraphics[width=3in]{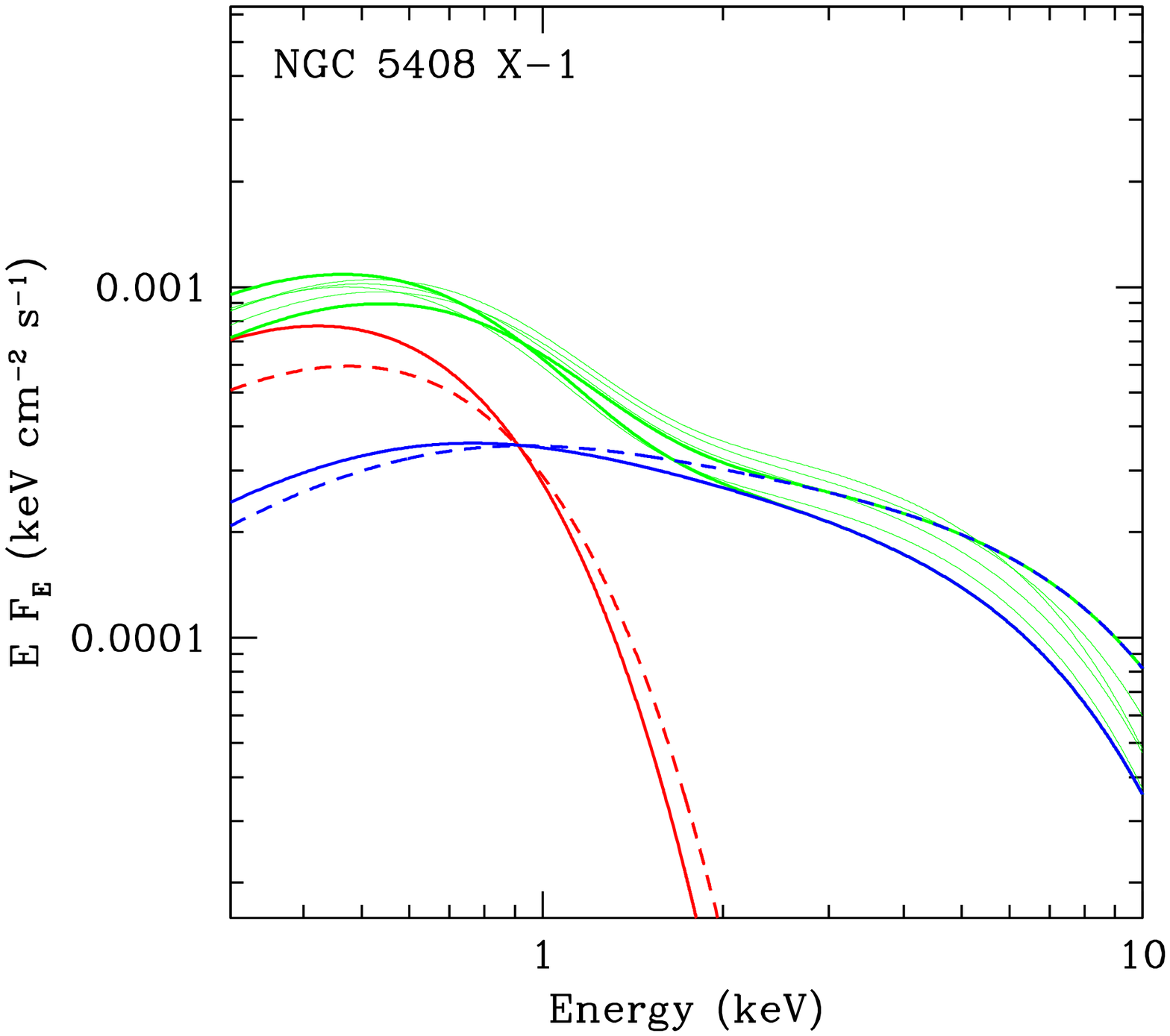}}\quad
\subfigure{\includegraphics[width=3in]{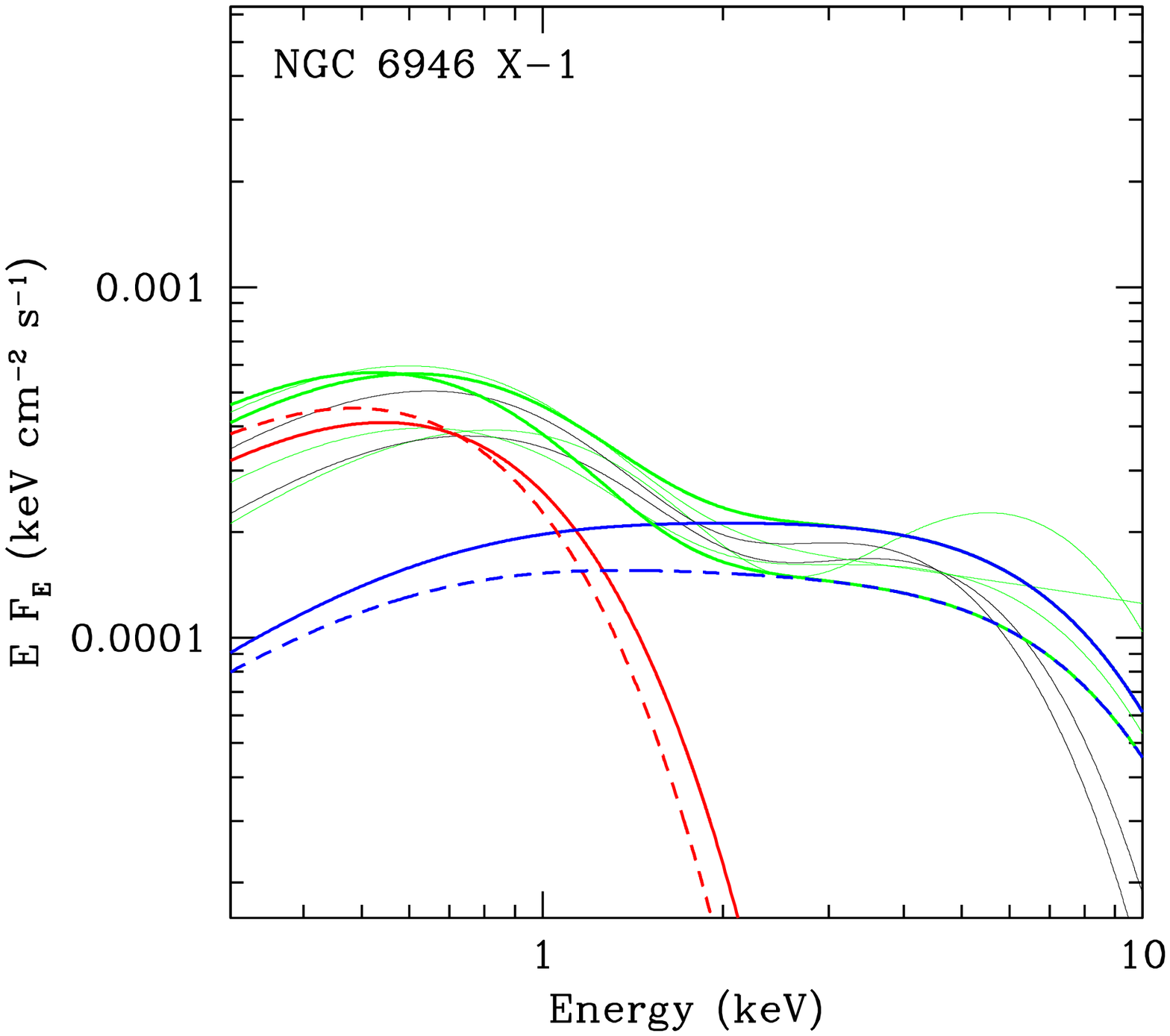} }}\\
\mbox{\subfigure{\includegraphics[width=3in]{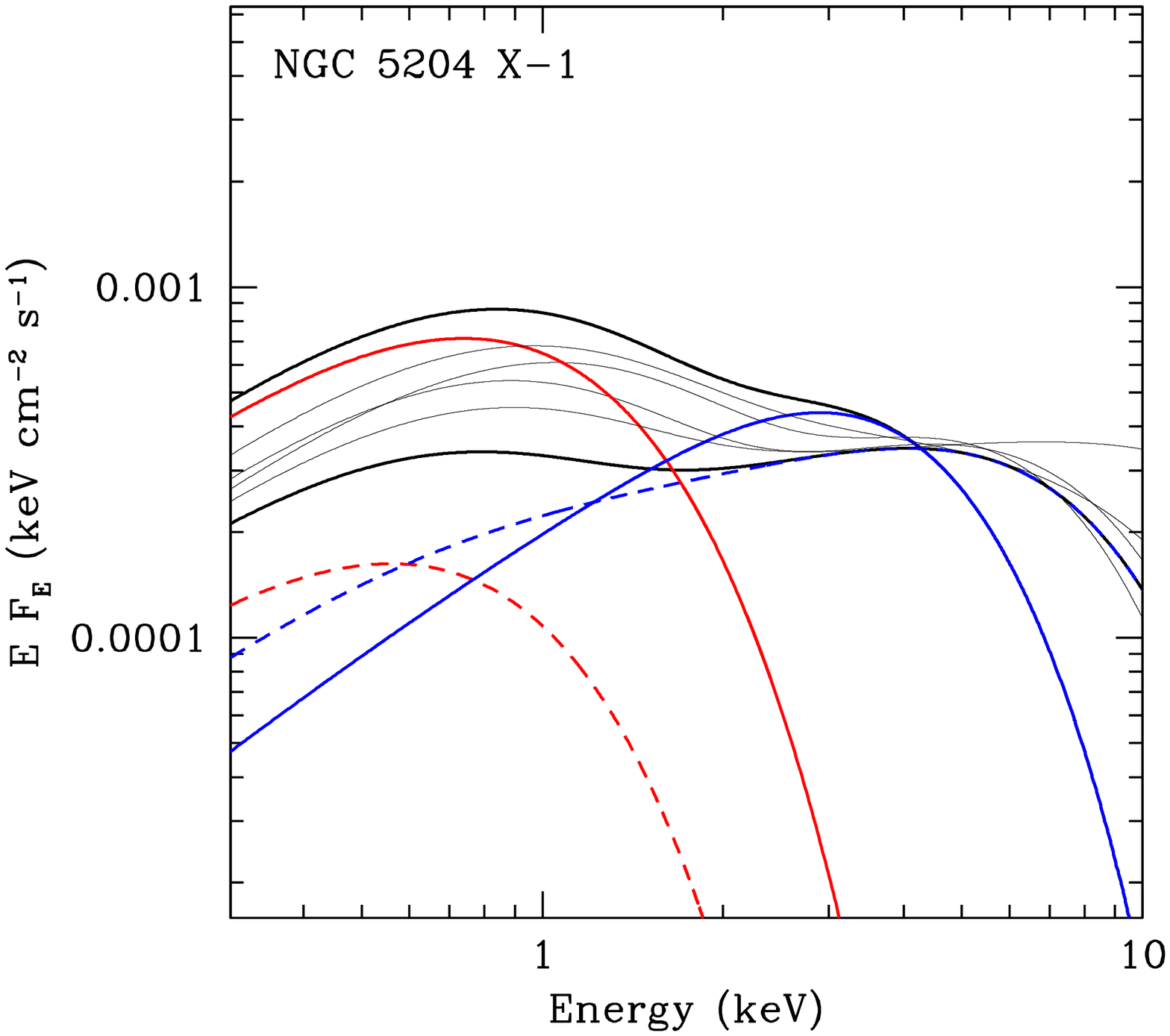}}\quad
\subfigure{\includegraphics[width=3in]{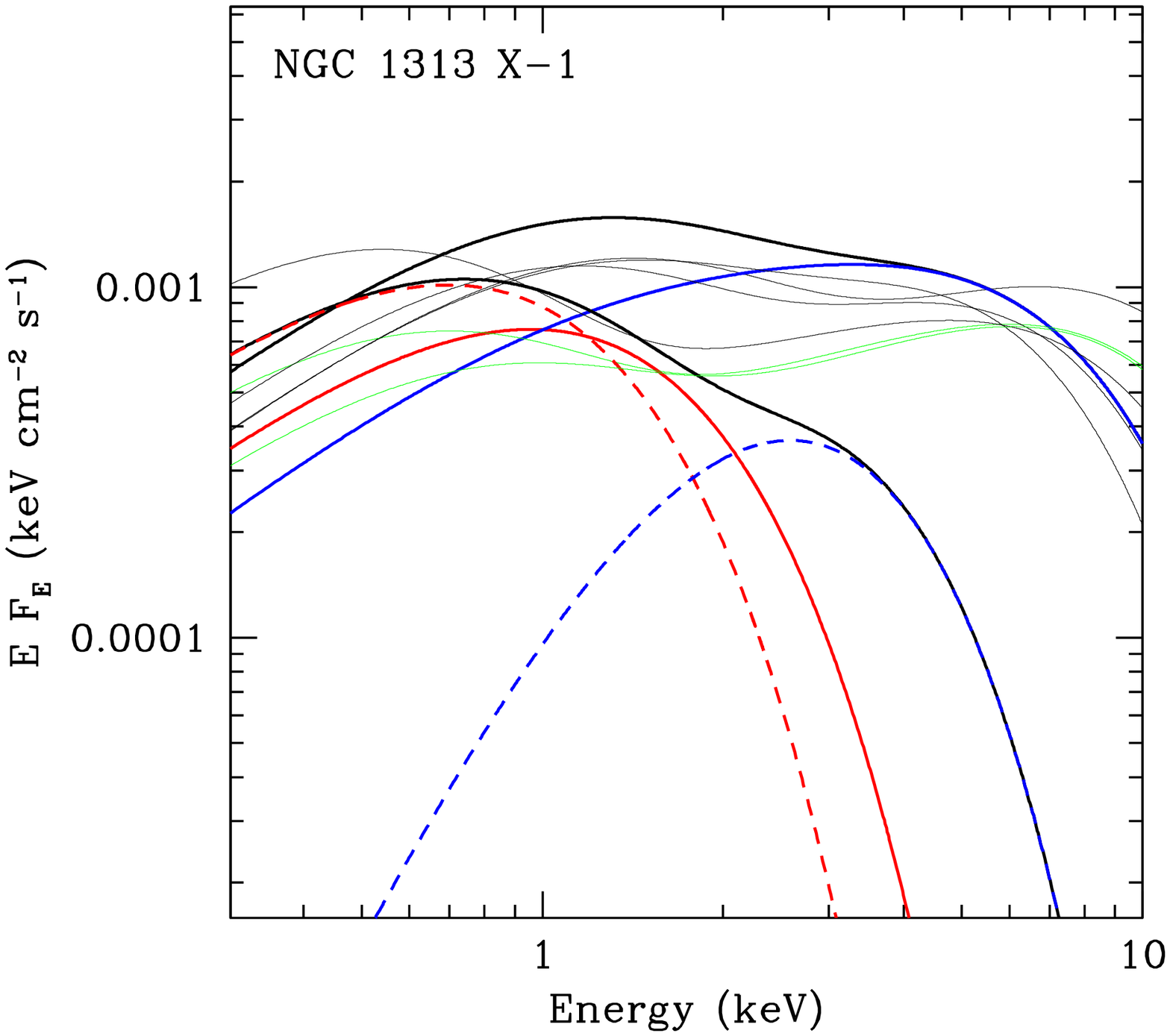} }}
\caption{Best-fitting continuum models (for model parameters see
  Tables 2 \& 3) for the ULXs as labeled. For those where we have
  significant variability ($\ge$ 3 sigma) the spectral models are shown in
  green. For the brightest and dimmest spectra (in unabsorbed
  0.3-10~keV flux, indicated by thicker solid lines) we also plot the
  model components, {\sc diskbb}: red solid/dashed (bright/faint)
  lines and {\sc nthcomp}: blue solid/dashed (bright/faint)
  lines. This allows for crude inspection of how the ULX has evolved
  from dimmest to brightest (although care must be taken as this may
  be misleading in certain cases: see NGC 1313 X-1 for an example in the discussion).} \label{fig12}
\end{figure*}

\begin{figure*}
\centering
\mbox{\subfigure{\includegraphics[width=3in]{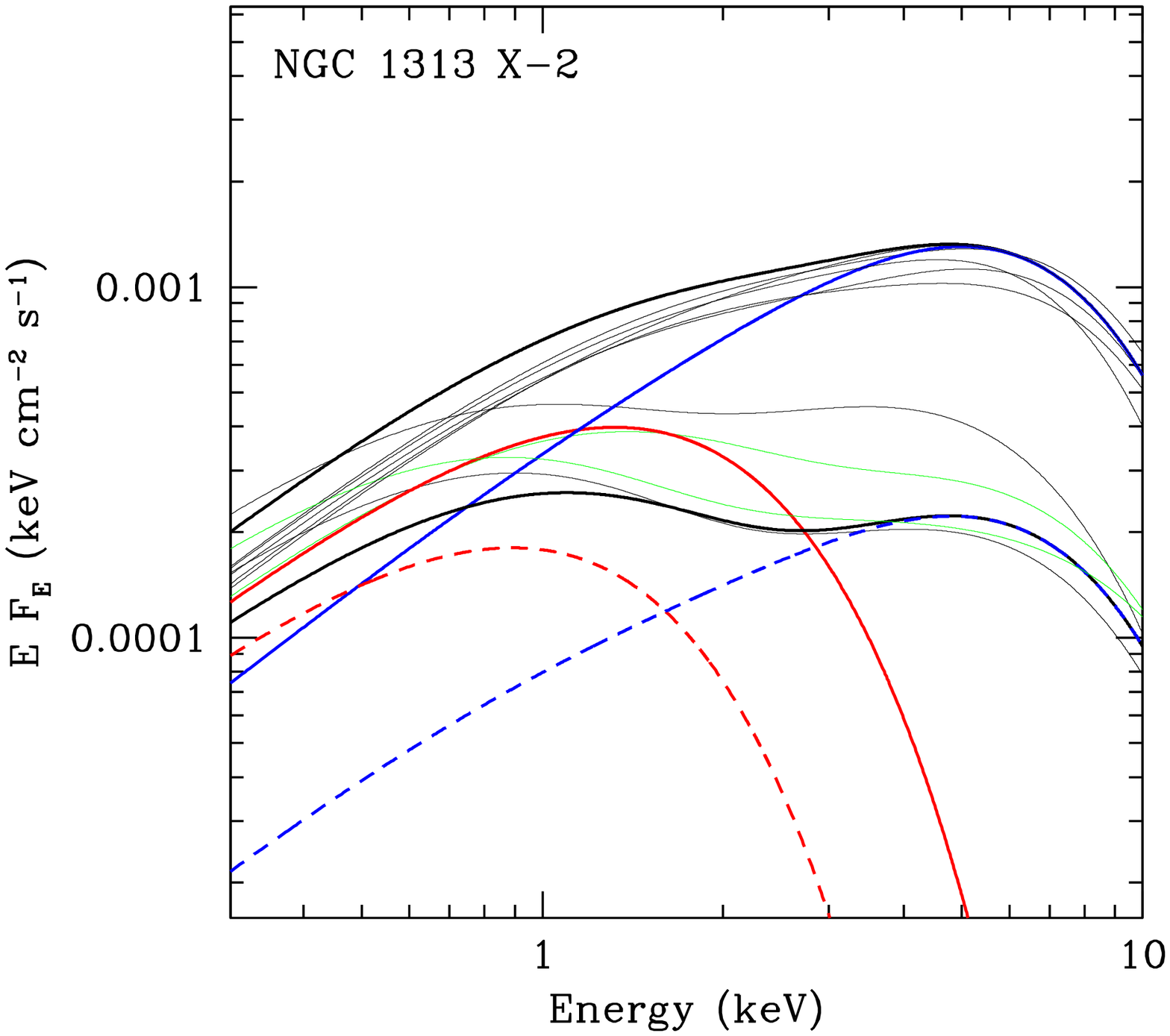}}\quad
\subfigure{\includegraphics[width=3in]{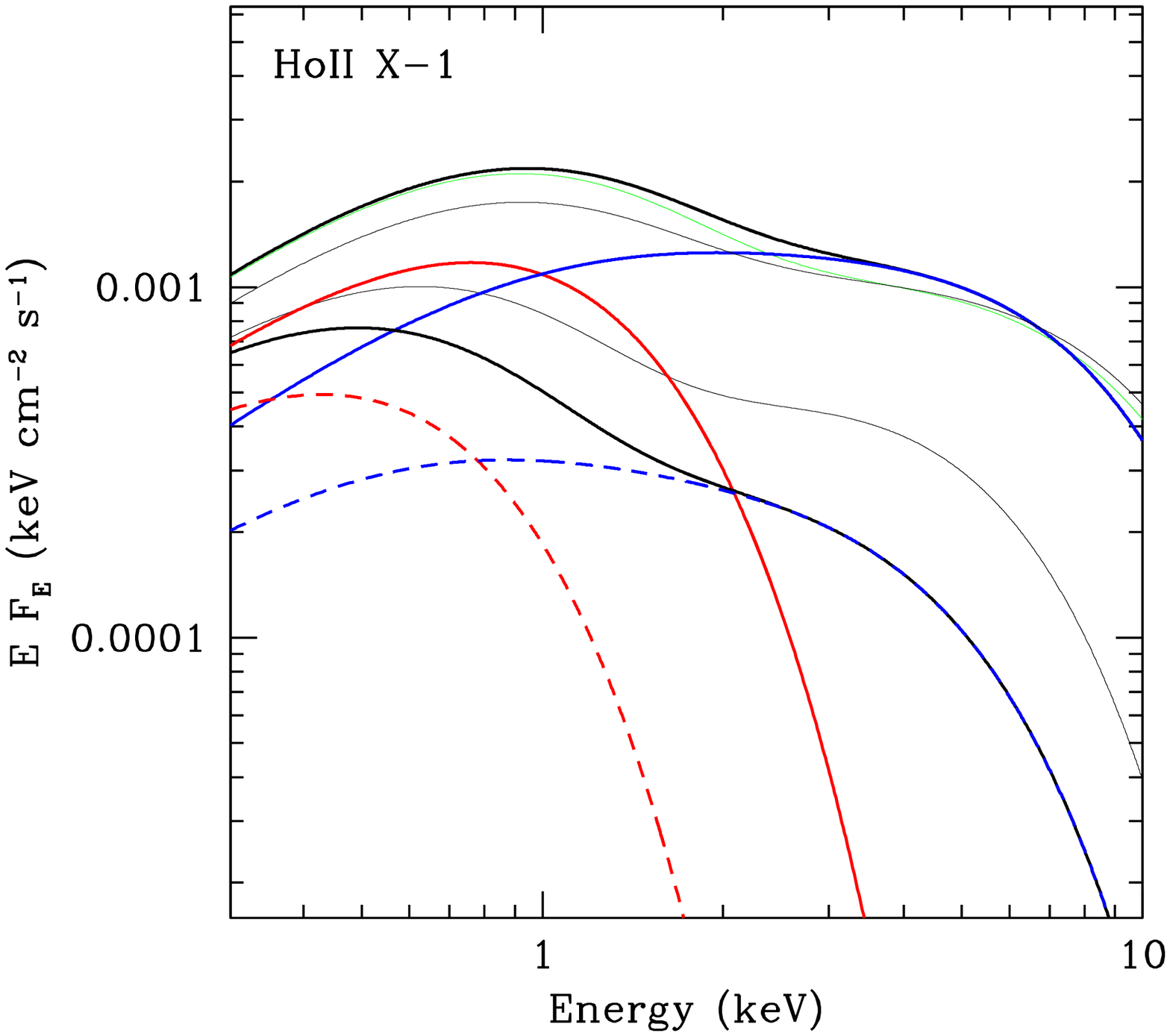} }}\\
\mbox{\subfigure{\includegraphics[width=3in]{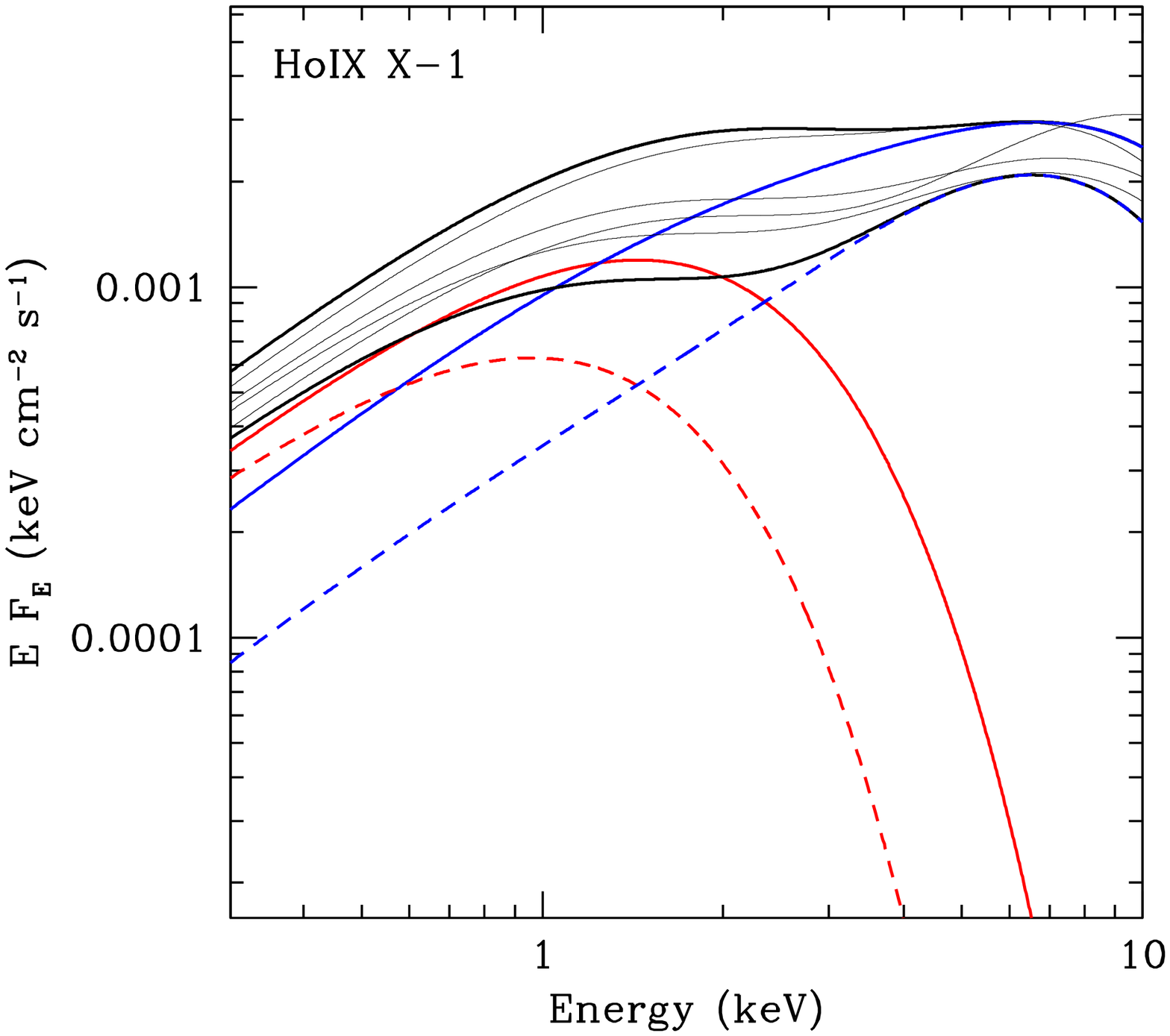}}\quad
\subfigure{\includegraphics[width=3in]{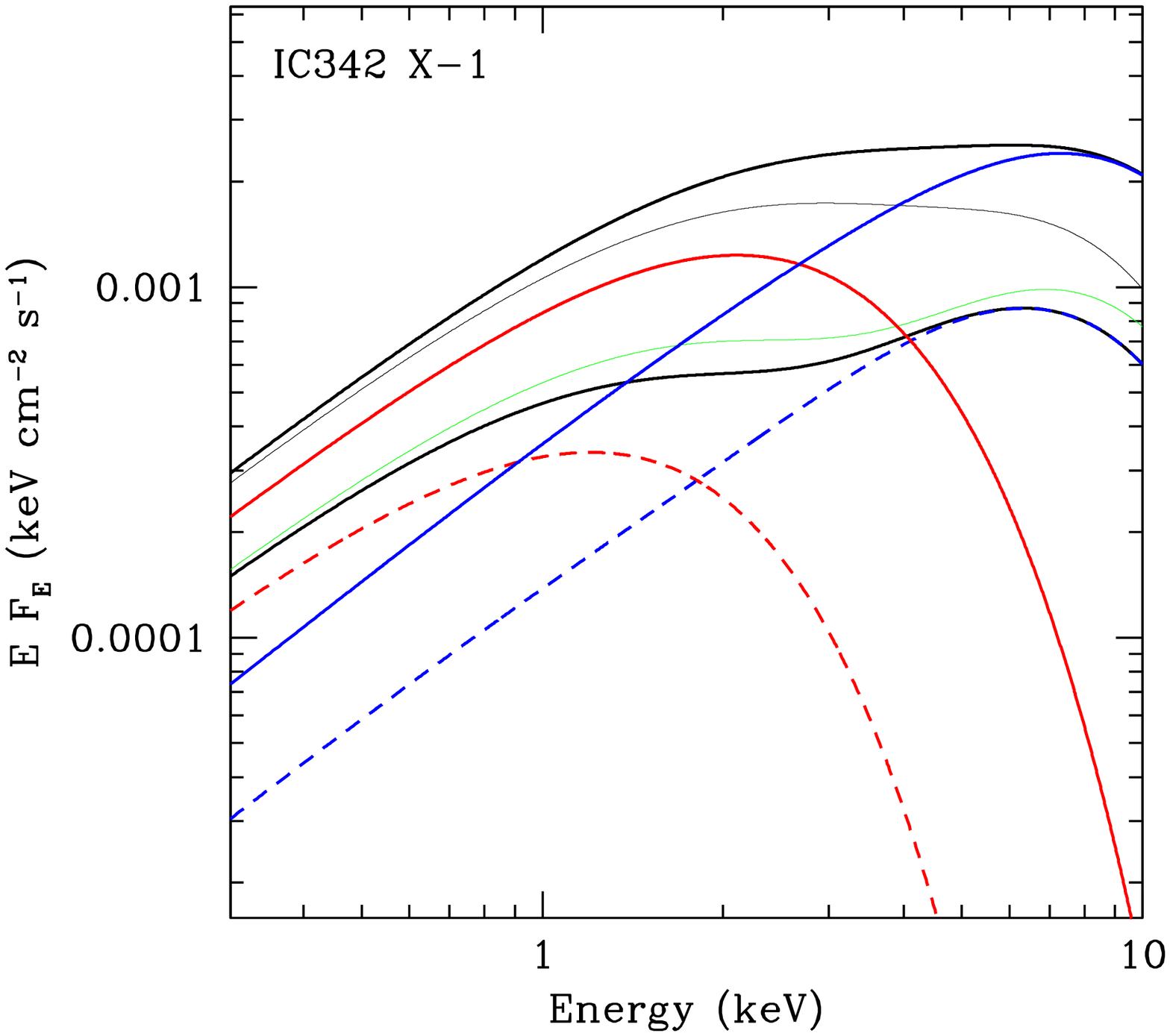} }}
\caption{As for Figure 5.} \label{fig12}
\end{figure*}

\subsection{Spectral analysis}

The 0.3-10~keV spectra of ULXs above $\sim$3$\times$10$^{39}$ erg
s$^{-1}$ generally favour a two component model description (Sutton et
al. 2013) with a high energy ($>$3 keV) break, distinctly unlike the
spectra observed from sub-Eddington accretion (Zdziarski et al. 1998;
Stobbart et al. 2006; Gladstone et al. 2009; Bachetti et al. 2013). In
this section, we model the ULX spectra assuming the soft emission
originates in the wind and hard emission from the inner, distorted disc
with down-scattering likely broadening the emission towards energies
below the peak.

Although advection is likely to be important for the wind at large
$\dot{m}_{\rm 0}$, we note that even the highest quality datasets do
not {\it presently} favour describing the soft emission with a model
for an advection dominated disc over a simple thin disc (though see section 5). As a result,
when fitting the data in {\sc xspec v 12.8} (Arnaud 1996), we use a
quasi-thermal, multi-colour disc blackbody ({\sc diskbb}: Mitsuda et
al. 1984) to account for emission from the wind (i.e. the emission
from $R_{ph,in}$ to $R_{\rm sph}$, e.g. Kajava \& Poutanen 2009) and a
broad model to describe the emission from the inner disc and its
down-scattered component. {\sc nthcomp} is an appropriate model for
the latter as it can provide a variety of complex continuum shapes
(Zycki, Done \& Smith 1999) whilst providing some useful and readily
identifiable `physical' parameters (e.g. the photon-index of the
spectrum, $\Gamma$, and high energy rollover: 2-3~$kT_{\rm e}$)
although we have to be careful in how we interpret these in the sense
of the physical model. We note that Walton et al. (2014; 2014b) show that the
high energy tail (in at least Ho IX X-1 and Ho II X-1) following the rollover is not well described by a
Wien tail. As this is out of our bandpass it will not affect our
spectral modelling (as our model hinges on the characteristic
temperatures) although we speculate that this broadening may result
from the heavily distorted inner disc emission (see also Tao \& Blaes
2013).

\begin{figure}
\begin{center}
\begin{tabular}{l}
 \epsfxsize=8cm \epsfbox{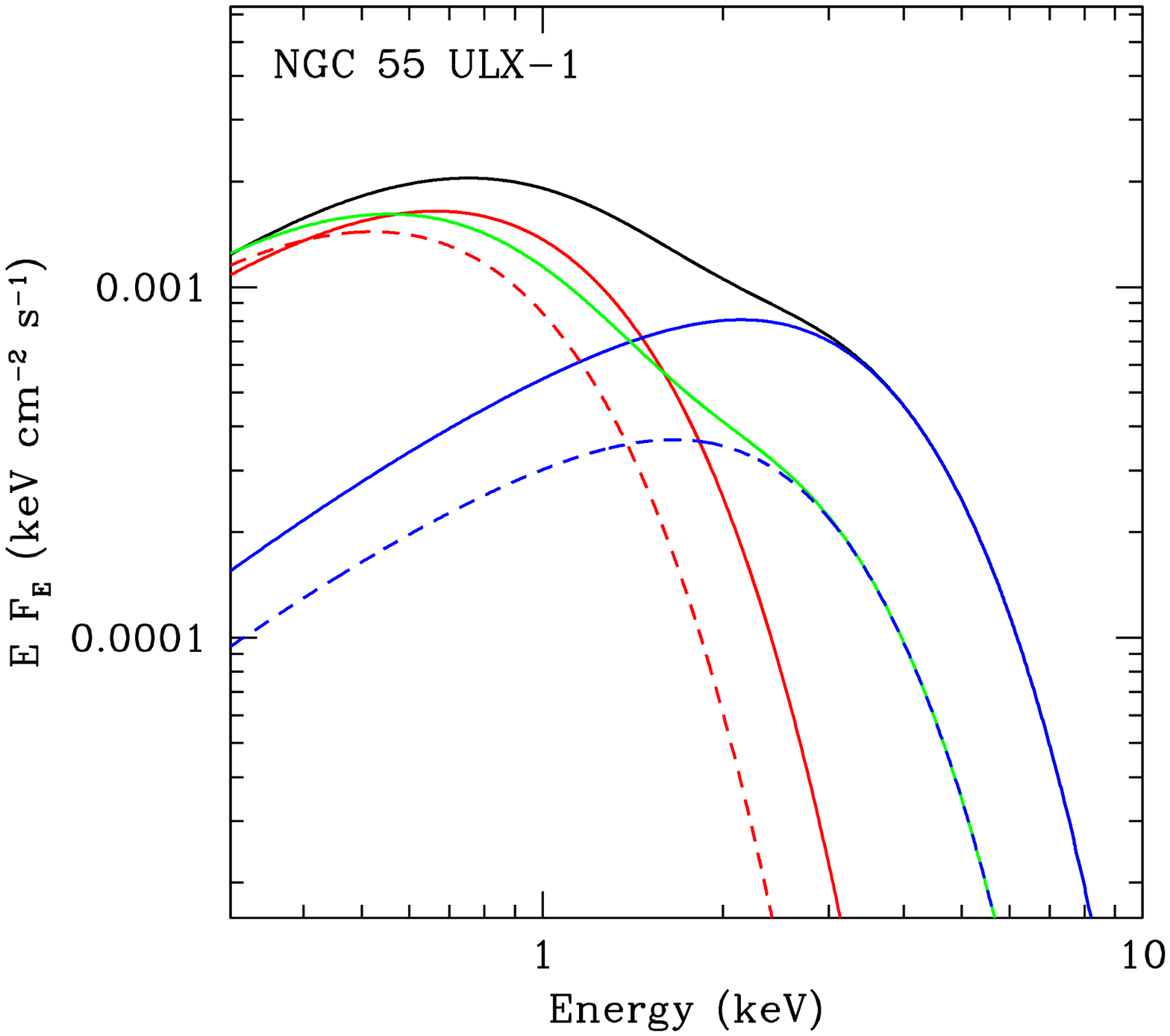}
\end{tabular}
\end{center}
\vspace{-0.5cm}
\caption{As for Figure 5.}
\end{figure}

We combine these emission components with a model for neutral
absorption ({\sc tbabs}) using appropriate abundance tables (Wilms,
Allen\& McCray 2000) with the lower limit set at the Galactic column
in the line-of-sight to each host galaxy (Dickey \& Lockmann 1999). We fit
this simple convolved model ({\sc tbabs*(diskbb+nthcomp)}) to all of
the ULX datasets, keeping the high energy component within a region of
parameter space such that the high energy rollover (which here refers
to the inner disc peak temperature, i.e. $2-3 kT_{\rm e} \approx kT_{\rm in}$) is in the observable bandpass (as
seen in Stobbart et al. 2006; Gladstone et al. 2009). This is
preferable to allowing the model parameters to roam freely which can
often lead to the high energy rollover being out of the bandpass,
which in light of our model would be incorrect (i.e. we do not expect
a further break in the spectrum, as indicated by {\it NuSTAR}
observations: Bachetti et al. 2013; Walton et al. 2014).  As a result,
the index ($\Gamma$) is associated with the rise to the peak of the
high energy emission and
allows for a contribution from down-scattering.

In using the {\sc nthcomp} component to model emission from the hot
inner disc and down-scattered component, we fix the seed photon
temperature to be that of the wind to provide a realistic (albeit conservative) constraint:
by energy balance we expect the down-scattering to lead to
temperatures $\apgt$ $T_{\rm sph}$. For the highest quality (or
softest) datasets we note that the models favour the wind emission
component dominating at soft energies ($\aplt$ 1keV) over the hard
component. However, due to poor data quality in several datasets or
where the data does not obviously show a two-component structure, it
becomes necessary to force the presence of a soft component dominating
below $\aplt$ 1keV rather than allowing a fit with only a single
(broad) component. As a consequence, the errors on the parameters in
these models (denoted by a $^{*}$ or $^{**}$ in tables 2 \& 3) are
strictly only an approximation as they are found by fixing the
temperatures ($kT_{\rm d}$ and $kT_{\rm e}$) and/or the slope
($\Gamma$) for each component in turn (and rarely by fixing the
normalisation of the {\sc diskbb} component).

Across all observations we generally obtain a fairly hard ($\Gamma <$
2.4) photon index and a rollover in the spectrum $>$~3~keV (as reported
in Gladstone et al. 2009; Bachetti et al. 2013). The best-fitting
parameters of interest from model-fitting are presented in Tables 2 \&
3 (with the total absorbed luminosities also provided in Table 1) and
spectral plots in Figures 5, 6 \& 7.

In order to make comparisons to our predictions for the
spectral-timing evolution of ULXs as a population, we obtain the
de-absorbed fluxes in two energy bands: 0.3 - 1~keV and 1 - 10~keV
respectively by including a {\sc cflux} component in the model fitting
({\sc tbabs*cflux*(diskbb+nthcomp)}); these values are given alongside
the model parameters in Tables 2 \& 3. We then determine the `colour'
from the ratio of hard to soft fluxes and determine the error on the
colour from propagation of errors. Together with the fractional rms
(see next section), we can subsequently assess the evolutionary
pattern of ULXs in spectral hardness and variability.

\subsection{Timing analysis}

We obtain the PDS (normalised to be in [$\sigma$/mean]$^{2}$ units) by
fast-Fourier transforming segments of background subtracted,
0.3-10~keV, time series of length 1200~s and taking the average of the
resulting periodograms (see van der Klis 1989 for a review). Short
periods ($<$ 10~s) of instrumental dropouts are then corrected by
linearly interpolating between the points either side of the gap. By
integrating the (Poisson noise subtracted) power over a given
frequency range and taking the square root we are naturally left with
the fractional rms ($F_{\rm var}$: see Edelson et al. 2002). This is a
more robust method of measuring $F_{\rm var}$ than by obtaining the
variance of the lightcurve in the time domain as multiple measurements
of the variance are taken and the true distribution can be measured
(see Vaughan et al. 2003). We integrate the PDS over a frequency range
3 to 200 mHz - as this range encapsulates the broad-band noise
variability seen in most ULXs (Heil et al. 2009) and ensures adequate
statistics in the majority of cases - with the resulting values
presented in Tables 2 \& 3. We note that, although QPOs are seen in two of the sources presented here (NGC 5408 X-1 and NGC 6946
X-1) and lie within the frequency range over which our rms
is calculated, in neither case will the variability of the QPO
dominate over the broadband noise (see e.g. Strohmayer \& Mushotzky 2009) and so our values fairly
reflect the underlying variability. 

Constraining a value for $F_{\rm var}$
depends on both the level of Poisson noise (set to be 2/mean count
rate in our normalisation) and the length of an observation; for those
observations where the former is high or the latter short we may not
be able to directly constrain the presence of variability. For observations with unconstrained fractional rms, upper limits were found by performing lightcurve simulations with an underlying PSD shape equivalent to that of a highly variable observation of NGC 6946 X-1 (Obs.ID: 0691570101), determined by fitting Lorentzians to the PDS. Simulated lightcurves were generated with the same count rates and duration as the real observations. In order to place upper limits on the power (with a similar shape to that of NGC 6946 X-1), hidden within each PDS, the input model was rescaled at 1\% intervals from 20-100\% of the normalisation of the original PDS. For each rescaled input model, 200 lightcurves were generated and the average rms and its error calculated; the upper limit was then taken to be the point where the rms was just significant at a 3 sigma level and is given in Tables 2 \& 3.

We plot the power (i.e. fractional rms squared) for each
observation (where constrained) against the spectral hardness in
Figure 8. In plotting these together we have assumed that the
variability timescales between sources are the same (reasonable in
light of expected viscous timescales: see section 2) and that we have
a homogeneous population of compact objects. Although
differences in the mass transfer rate will lead to differences in the
dampening of variability (as $\gamma$ in Method 2 is expected to be a
function of $\dot{m}_{\rm 0}$), by including a large enough sample of
sources (and observations) we can start to average over this
difference.

Although there is considerable scatter, the general trend appears to
be broadly consistent with the predicted shape of spectral hardness vs power shown
in Figure 4. The amount of scatter is unsurprising given the range of possible evolutionary tracks (driven by the effect of precession and the wind tending towards homogeneity at large $\dot{m}_{\rm 0}$: section 3) and that there will be a distribution of masses. We note that should this association be correct,
the positive gradient branch would appear to be underpopulated; this
is likely due to selection effects as these will be under-luminous and
may not qualify as bright `ULXs' in the traditional sense (see section
3).

To estimate the significance of a correlation on the negative slope we
exclude NGC 55 ULX-1 - as the spectrum is significantly softer than
any other ULX (see next section for more details) - and perform a Kendall's rank coefficient test on the
data binned into 5 even logarithmic spectral hardness bins (which does
not account for the errors on the data), however, this is only
marginally (2-sigma) significant. We can however, attempt to rule out a positive slope, expected
should we be observing a population of IMBHs (which we assume behave
as BHBs at similar Eddington ratios: Belloni 2010; Mu{\~n}oz-Darias et al. 2011). We
determine the gradient of the slope through a Least-squares fit to the
un-binned datapoints (using average symmetrical errors),
constraining the gradient to be negative at $>$ 3-sigma ($<$ -0.14 at 3-sigma). This would seem to rule
out the presence of a positive slope although we caution that the
fit is heavily influenced by the 5 points at high hardness and further
data would be useful to confirm this.

Whilst the hardness-variability trend is consistent with the
expectations of our model, we proceed to investigate individual source
behaviour to further test whether the observations match (or can be
broadly explained by) the detailed predictions discussed in section 3.

\begin{figure*}
\begin{center}
\begin{tabular}{l}
 \epsfxsize=14cm \epsfbox{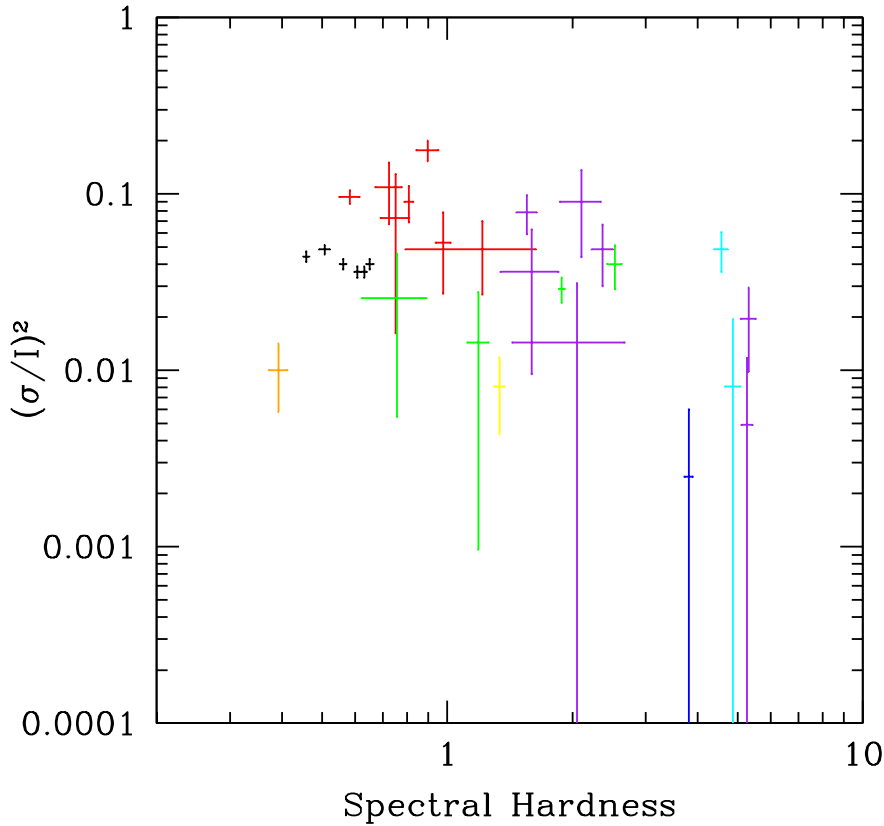}
\end{tabular}
\end{center}
\vspace{-1cm}
\caption{Spectral hardness (from the ratio of unabsorbed 1-10/0.3-1
  keV fluxes) versus power (fractional rms squared) in log
  space (with 1 sigma errors). The
  colour scheme corresponding to the different sources is as follows:
  black: NGC~5408 X-1, red: NGC 6946 X-1, purple: NGC 1313 X-2,
  orange: NGC~55 ULX-1, green: NGC 1313 X-1, cyan: IC 342 X-1, blue:
  Ho IX X-1, yellow: Ho II X-1. This appears similar in overall shape
  to our predicted evolution shown in Figure 4.}
\end{figure*}

\section{Energy-dependent variability}

A clear prediction of our spectral-variability model is
that at small $\theta$, variability should be
present mostly on long timescales (see section 2) and should be a complex combination of the variability directly from the disc (assuming that the disc is not thin: Churazov et a. 2001),  
Compton down-scattered emission from the scattering surface of the wind and changing $f_{\rm col}$. At
intermediate $\theta$ we should see a slightly different energy
dependence of the variability (as obscuration events are expected to dominate where $m_{\rm 0}$ is not too large) with the same spectrum as
the inner disc emission (which arrives to the observer down-scattered).

The energy dependence of the variability in NGC 5408 X-1 has already
been shown in Middleton et al. (2011) with the fractional rms
increasing with energy. However, it is clear that those spectrally hard sources where
the variability is predicted to be created by scattering {\it into}
the line of sight (i.e. those at small $\theta$) all have low rms and
so we cannot investigate the shape of the rms spectrum. We can improve on
this situation by extracting the cross spectrum
(Nowak et al. 1999) and selecting the linearly correlated/coherent
variability relative to a reference band with high signal-to-noise
rms. Normalising by the excess variance in the reference band and
plotting this against energy gives the covariance spectrum (Wilkinson
\& Uttley 2009) where the removal of uncorrelated Poisson noise
significantly reduces the sizes of the errors on each data-point (of
variance) in each energy bin. We plot the covariance (extracted in the
time domain) relative to the 1.5 - 3~keV band, for the best
constrained single observation of each source (with the exception of
IC 342 X-1, NGC 1313 X-2 and NGC 5204 X-1 due to insufficient data quality) in Figures
9, 10 and 11 over two timescales: long (0.9 - 3 mHz) and short (3 -
200 mHz). The short timescale variability corresponds to the fractional rms values reported in Tables 2 and 3 and we expect to include the contributions from both methods of generating variability discussed in section 2.
The covariance spectra interrogated on long timescales allow for a comparison (where variability on both timescales can be constrained) but also allows us to investigate the correlated spectral variability for sources with low rms on the shorter timescales. We plot these unfolded through, and plotted alongside, the
de-absorbed, best-fitting time-averaged model for each observation to
allow ease of comparison.

  \begin{figure*}
\centering
\mbox{\subfigure{\includegraphics[width=3in]{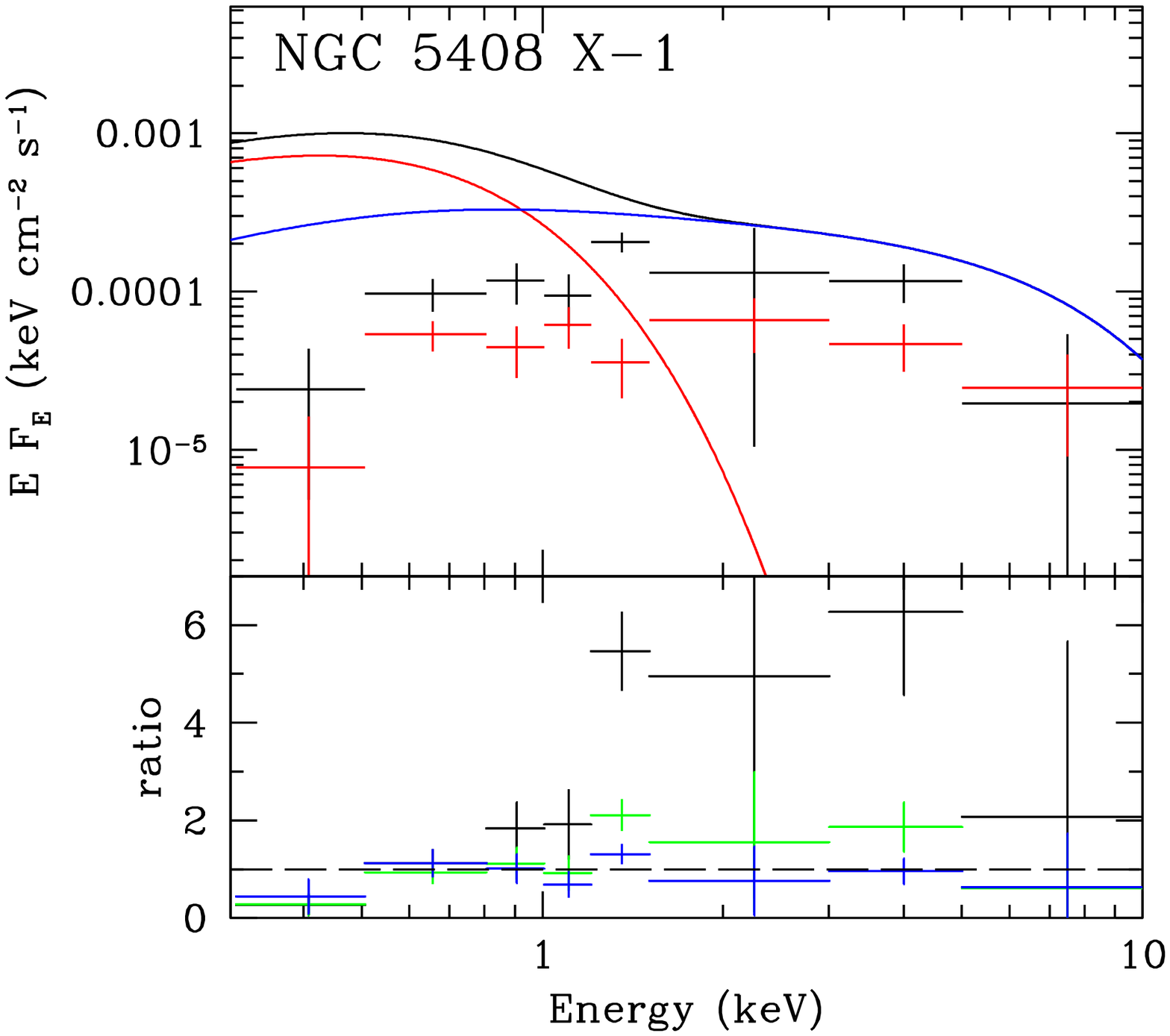}}\quad
\subfigure{\includegraphics[width=3in]{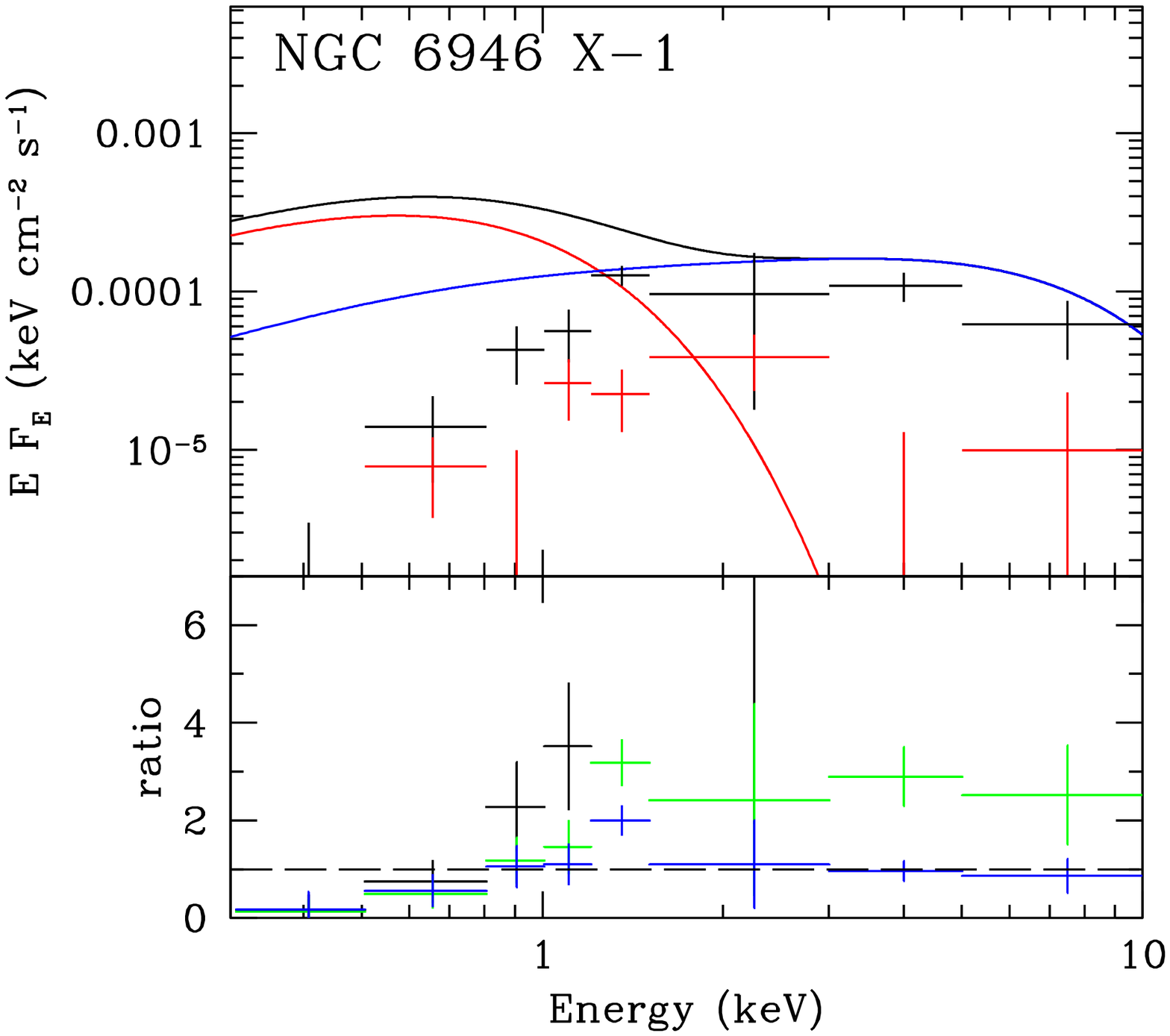} }}\\
\caption{Upper panels: Covariance spectra relative to the 1.5-3~keV band, over two
  timescales: long (red: 0.9-3mHz) and short (black: 3-200mHz) for NGC
  5408 X-1 (Obs.ID: 0500750101) and NGC 6946 X-1 (Obs.ID: 0691570101) plotted with
  (and unfolded though) their best-fitting, de-absorbed, time-averaged
  spectral model. Lower panels: ratio plots from fitting the covariance spectra with the time-averaged, de-absorbed model with the relative amount of {\sc diskbb} to {\sc nthcomp} fixed (black), free (green) and with the {\sc nthcomp} component only, with free model parameters (blue; note that $kT_{\rm e}$ is fixed to its best fitting value in the time-averaged spectral fitting). When the normalisations are free, the {\sc diskbb} component is not required; clearly the variability is therefore associated with the hard component. The ratio plots also show that, whilst the variability spectrum is
  a good match to the high energy component above $\approx$1~keV, at
  softer energies the model over-predicts the data as expected from
  our model assumptions (section 3).}
\label{fig:l}
\end{figure*}

\begin{figure*}
\centering
\mbox{\subfigure{\includegraphics[width=3in]{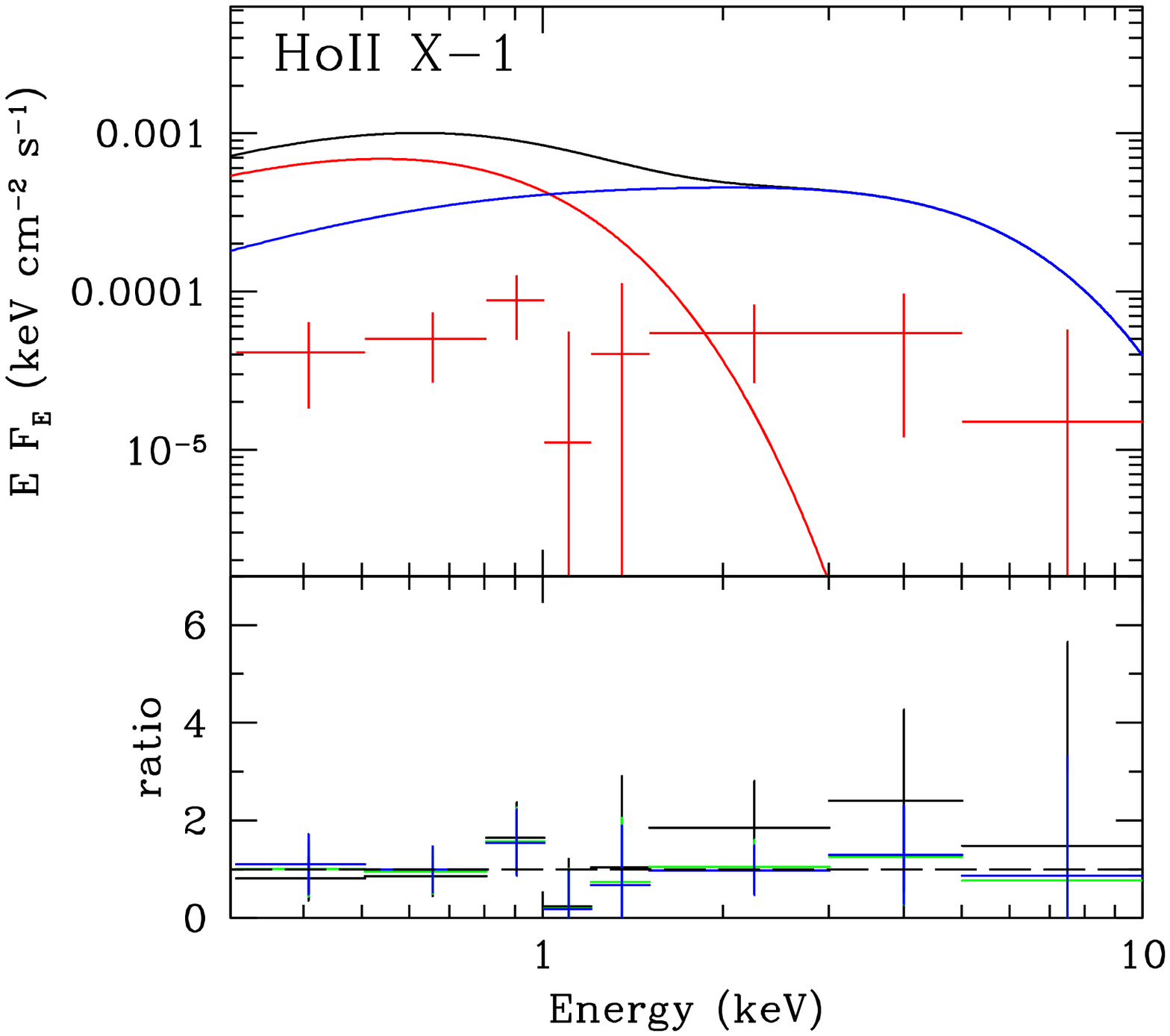}}\quad
\subfigure{\includegraphics[width=3in]{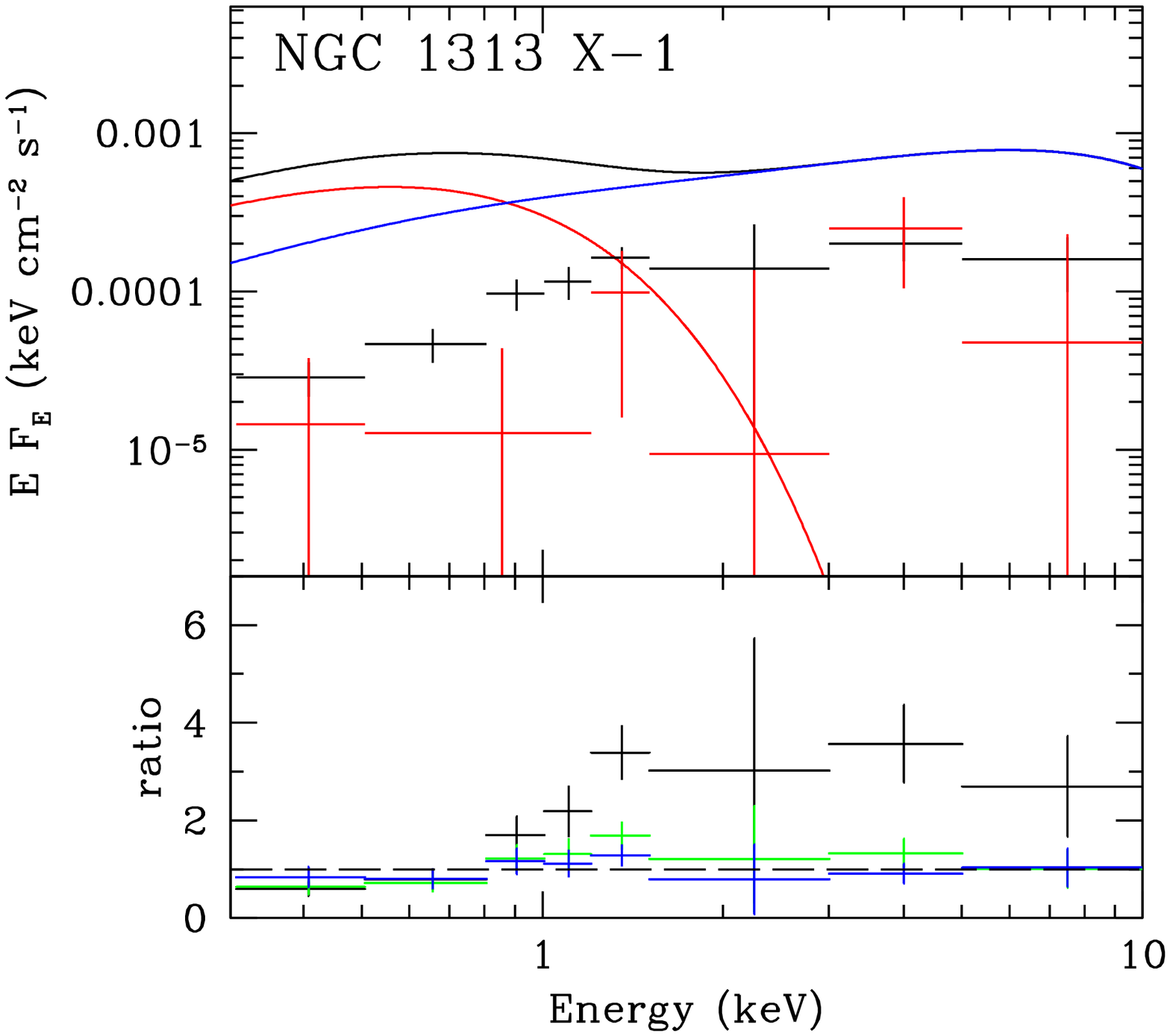} }}\\
\caption{As for Figure 9, covariance spectra for Ho II X-1 (Obs.ID: 0561580401), which can only be extracted on long timescales, and NGC 1313 X-1 (Obs.ID: 0405090101).}
\label{fig:l}
\end{figure*}

\begin{figure}
\centering\includegraphics[width=3in]{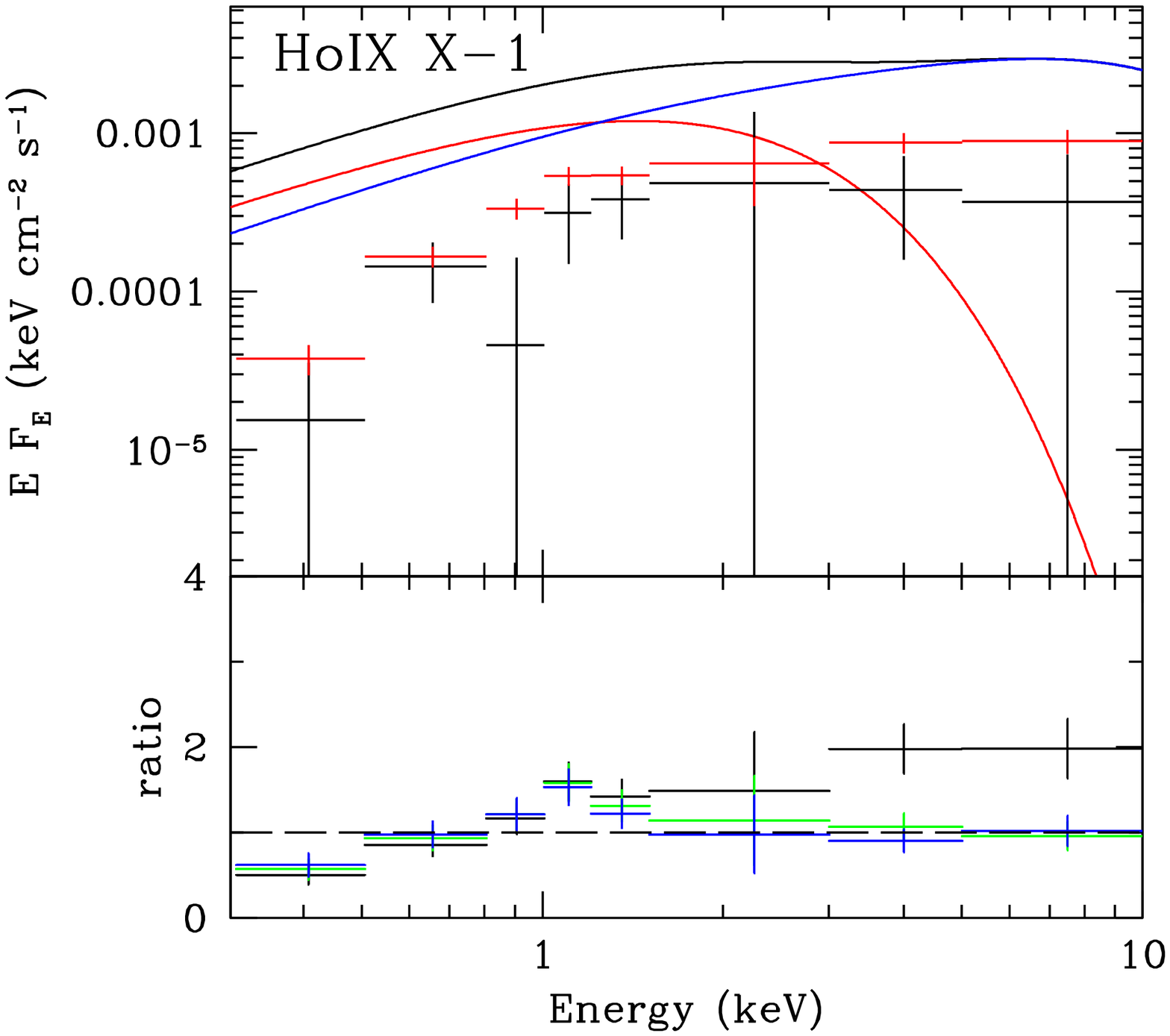}
\caption{As for Figure 9, covariance spectrum of Ho IX X-1 (Obs.ID: 0657801801).}
\label{fig:l}
\end{figure}

In the case of the softer sources (NGC 5408 X-1, NGC 6946 X-1 and to a
lesser extent, Ho II X-1: Figure 9 \& 10), we see a shape consistent
with the variability originating in the hard component {\it only}, consistent with a model where the variability on short and long timescales at
such inclinations is dominated by obscuration of the high energy
emission (Method 1). To demonstrate this more clearly, we fit the short-timescale covariance data (long timescales in the case of Ho II X-1) with the de-absorbed, time averaged model where the component normalisations relative to one-another are initially fixed (so that there is the same proportion of {\sc diskbb} to {\sc nthcomp} as in the time-averaged, best-fitting model); this leads to large residuals (plotted as a ratio in the lower panels of Figures 9 \& 10 ). We then proceed to free the normalisations, finding that the residuals improve and the {\sc diskbb} normalisation tends to zero.  It is apparent that, whilst the variability
is generally well matched by our choice of hard component model above 1 keV, the
spectral modelling may deviate somewhat at the very softest energies (most readily seen for NGC 6946 X-1). This is
not unexpected; we forced the lowest temperature of the {\sc nthcomp}
component to be that of the wind via energy balance, however, this
assumes the most extreme scenario where the down-scattering occurs by
the coolest wind material at the largest radii. Instead, for sources
at more moderate or low inclinations, the temperature of the wind
plasma intercepting the hard photons is expected to be somewhat
higher, leading to the low energy peak (of the high energy component)
moving to higher energies (or in other words $kT_{\rm in} > kT_{\rm
  d}$ in our model). As we are using thermal components in our
spectral fitting rather than simple power-laws, the effect on the
spectral hardness will only be small (as the spectrum is downturning
to meet the peak of the seed photon distribution: Figures 5, 6 \& 7)
and so our spectral-variability trend (Figure 8) remains robust. As an
interesting aside, as the hard component rolls over at lower
energies/faster than we have accounted for, it seems likely that the
soft component could be of a different shape to that determined from the time-averaged spectra (which we will investigate elsewhere). In order to obtain the best description of the covariance spectra, we allow the parameters of the {\sc nthcomp} component to be free with the exception of the peak temperature, $kT_{\rm e}$, set to the best-fitting value in the time-averaged spectrum (Tables 2 \& 3) to keep the rollover within the  bandpass. We also restrict the lower limit of $kT_{\rm d}$, to that of its time-averaged, best-fitting value as it is clear that the true temperature likely resides at or above this value. Fitting with the {\sc nthcomp} component only, provides the greatest improvement as shown in the lower panels of Figures 9 \& 10. The best-fitting parameters and their errors are given in Table 4, although we caution that these values should only be seen a representative of the spectral shape rather than a physical description. It is interesting to note that, in the case of NGC 6946 X-1, even this component alone does not appear to account for the shape of the covariance at the softest energies.

\begin{table*}
\begin{center}
\begin{minipage}{110mm}
\bigskip
\caption{{\sc nthcomp} best-fitting parameters for the covariance spectra}
\begin{tabular}{lc|c|c|c|c|}
  \hline

Source & Timescale & $kT_{\rm d}$ (keV) & $\Gamma$ & norm ($\times$10$^{-4}$)\\
 \hline
NGC 5408 X-1 & short &  0.88$^{+0.16}_{-0.38}$ &  $>$ 2.36   & 1.2 $\pm$ 0.2  \\
NGC 5408 X-1 & long &  $<$ 1.14 &      $>$ 1.74   &   0.4 $\pm$ 0.1\\
NGC 6946 X-1 & short & 1.25$^{+0.52}_{-0.62}$  &   $>$ 1.69   & 0.5 $\pm$ 0.1   \\
NGC 6946 X-1 & long &  0.94$^{+0.74}_{-0.70}$ &   $>$ 1.61  & 0.1  $\pm$ 0.1 \\
HoII X-1 & long &  $<$ 1.02 &  $>$ 1.56    & 0.6 $\pm$ 0.3  \\
NGC 1313 X-1 & short & 1.20$^{+0.56}_{-0.64}$  &  $>$ 1.72   & 0.9 $\pm$ 0.1 \\ 
NGC 1313 X-1 & long & $>$ 0.24  & unconstrained   & $<$ 0.8\\
HoIX X-1 & short &  $<$ 3.38 &  $>$ 1.39  &   1.8$^{+0.9}_{-0.8}$   \\  
HoIX X-1 & long &  1.43 $^{+0.68}_{-0.65}$ & $>$ 1.71   &  3.1 $\pm$ 0.4   \\  
NGC 55 ULX-1 (Obs 1) & long & $<$ 1.34 &    1.46$^{+0.80}_{-0.25}$   & 0.4 $\pm$ 0.1  \\
NGC 55 ULX-1 (Obs 2) & long &   $<$ 0.96       &  $>$ 1.29  & 0.3 $\pm$ 0.1 \\
   \hline

\end{tabular}
Notes: {\sc nthcomp} model parameters (and their 90\% errors) for the best-fit to the covariance data shown in Figures 9-12.
\end{minipage} 

\end{center}
\end{table*}

The covariance spectra of Ho~IX~X-1 and NGC 1313 X-1 (Figures 10 \& 11) highlight the variable
component for the hardest ULXs which we have not been able to
previously study. Once again we show the residuals (as a ratio) from fitting the covariance (short timescales in the case of NGC 1313 X-1 and long timescales for Ho IX X-1) with the de-absorbed, time-averaged spectral model with a fixed proportion of normalisations, free normalisations and with the {\sc nthcomp} component alone (with free model parameters, with the exception of $kT_{\rm e}$). In the case of the highest data-quality, i.e. NGC 1313 X-1, we see that the best description of the covariance spectrum is once again the hard component on its own (i.e. the {\sc diskbb} normalisation tends to zero) with free model parameters (see Table 4). Based on our model, we may have expected the 
variability spectrum to have a contribution due to down-scattering below
that of the hard component with the addition of variability in the soft component
due to changing $f_{\rm col}$. Although the covariance spectrum
may appear slightly flatter than the de-absorbed, time-averaged model, consistent with expectation (appearing as an excess at $\sim$1-2~keV above the {\sc nthcomp} component in the residuals for Ho IX X-1 in Figure 11), the data quality is not yet sufficient for any unambiguous claim of such broadening.

Of particular interest for our model is the source NGC~55 ULX-1 as, due to its extremely soft spectrum,
this may sit on the positive branch of the hardness-variability plot
(Figures 4 \& 8). The source has been studied by Stobbart, Roberts \&
Warwick (2004) who suggested that the `dips' in the X-ray lightcurve
could be analogous to those seen in the class of dipping neutron star binaries
(i.e. obscuration by optically thick material in the outer
disc), implying a large (yet small enough that the source is X-ray bright) inclination to the source. Notably the dips get stronger with increasing energy and so also fit
into our model as a ULX seen at moderate-to-high
inclinations. Although the short timescale variability is
unconstrained in the first observation, on similarly long timescales
(0.03 - 3~mHz and 0.02-2~mHz respectively) both observations have
sufficient variability for the covariance spectrum to be extracted and
compared, and are shown in Figure 12. The second observation, which is
significantly softer than the first observation, is dimmer as we might
expect if the wind has become stronger or the system has
precessed. The covariance spectrum of the first observation is dominated by a hard component (as can be seen from the spectral residuals), with a peak temperature consistent with that of the time-averaged component, and appearing similar to those of other ULXs we have identified as moderately inclined (e.g. NGC 5408 X-1, NGC 6946 X-1). The second observation shows that the variability spectrum
is considerably flatter and the peak has shifted to lower energies
between observations. By fitting a simple blackbody ({\sc bbody} in {\sc xspec}) to
the covariance spectra we find that the peak has indeed shifted from
$kT_{\rm obs}$ = 0.71$^{+0.06}_{-0.05}$~keV to
0.28$^{+0.14}_{-0.07}$~keV and is no longer a reasonable match to the
temperature of the hard component within errors (Table 3). Under the reasonable assumption that the variability component is of the same physical origin, such a change is not expected in any model where the variable emission arrives to the observer directly (e.g. a disc-corona model as seen in BHBs). The nature of the evolution implies that the variable component of the emission has reached us after being down-scattered by cooler/more optically thick material than intercepts the hard emission. How this occurs is presently unclear but further observations of NGC 55 ULX-1 (and similar sources) may help place constraints on the required geometry.

\begin{figure}
\centering
\includegraphics[width=3.5in]{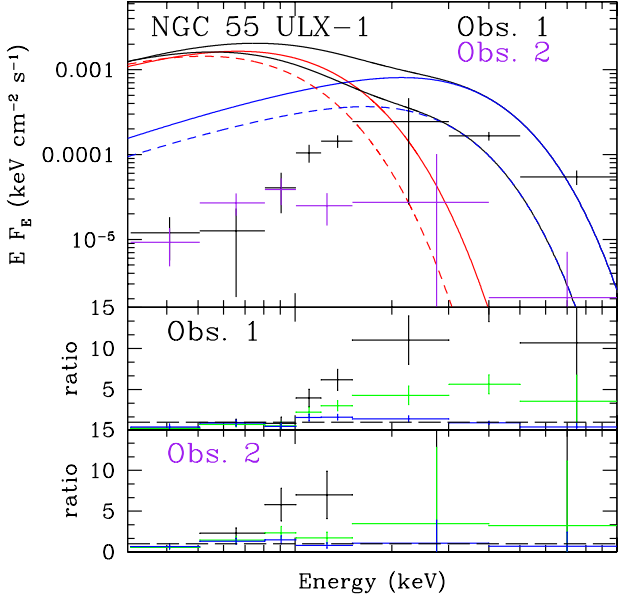} 
\caption{Covariance spectra of NGC 55 ULX-1 on long timescales (Obs.1: 0028740201, black: 0.03-3mHz; Obs.2: 0655050101, purple: 0.02-2mHz) which shows the decrease in peak temperature as the spectrum softens and dims, implying that we do not see the variable component of the emission directly. The lower panels show the residuals (as per the previous figures) for each observation in turn.}
\label{fig:l}
\end{figure}

\section{Discussion}

One of the leading models to explain ULX luminosities above
$\sim$3$\times$10$^{39}$ erg s$^{-1}$ is the formation of a powerful wind at
the spherization radius as a result of a high mass transfer rate from
the donor star (as seen in SS433, see Fabrika et al. 2004 for a
review). Radiative hydrodynamic instabilities are expected to be
present leading to a clumpy structure (Takeuchi et al. 2013;
2014). When combined with inflowing propagating mass accretion rate
fluctuations - now seen as the likely origin for the broad-band
variability characteristics of accreting sources (Lyubarskii 1997; Ingram
\& Done 2012; Uttley et al. 2005; Scaringi et al. 2012) - this leads to the
prediction of variability emerging directly (should the inner disc not be geometrically thin: Churazov et al. 2001) and/or via scattering events. 

In section 2 we presented two methods of imprinting variability which
vary as a function of inclination angle and mass accretion
rate. Combining these with the spectral predictions from P07, leads to
a series of observational spectral/timing predictions, as
outlined in section 3. These can be complicated by disc precession but
the overall trend we should expect across the population is one where
the harder sources, seen at small inclinations are generally less
variable than the soft sources seen at moderate inclinations. We also expect the variability to drop as the wind becomes
increasingly dominant and the source becomes softer. The predicted, inflected path (shown
in Figure 4), differs from that expected should the variability
originate in a high energy Compton-scattering plasma, as seen in BHBs,
where, crucially, the variability is greatest for the hardest source states (Belloni 2010; Mu{\~n}oz-Darias et
al. 2011). 

We combine spectral hardness, obtained from the
best-fitting models to the spectra of a sample of ULXs, with rms
values obtained from the PDS and plot these in Figure 8. There is
considerable scatter in the plot - expected in light of different evolutionary paths and a distribution of masses. 
However, we find a
significantly negative slope (when ignoring the softest datapoint -
which we have assumed sits on the positive slope of the predicted
trend) which is consistent with expectation and would disagree with
the expectation of IMBH accretion based on scaling BHBs to IMBH masses (Belloni 2010; Mu{\~n}oz-Darias et al. 2011). 

We also consider whether the variability could be caused by a variable, hot inner disc and stable wind, i.e. no contribution via Method 1. We note that, in the case of BHBs a thin disc can appear variable but only if illuminated by a variable photon flux (Wilkinson \& Uttley 2009), otherwise (under the assumption that variability becomes damped by the density of the disc) the disc would need to not be geometrically thin (Churazov et al. 2001), pointing to a situation unlike that expected for IMBHs. In Method 2 we consider that the propagating flux can emerge via scattering and/or from the inner disc directly. Assuming no contribution via Method 1 (i.e. obscuration by clumps), the implication would be that the soft sources are at a lower mass accretion rate so that the dampening due to mass loss is less severe (in effect more of the PDS is in our bandpass: see Figure 3 and section 2.2.3). Whilst we cannot rule this out, we still invoke a wind in this situation and so it seems highly likely that radiative hydrodynamic instabilities will occur leading to clumps and increased variability via Method 1 for these sources.

New models incorporating inhomogeneous accretion disc
emission (e.g. Dexter \& Quataert 2012) based on the photon bubble
instability (Gammie 1998; Begelman 2001) may also describe the X-ray
emission (Miller et al. 2014) and may offer a viable alternative to
distortion by radiatively driven winds. However, a key test of
any such model will be their ability to explain and incorporate the
changing variability properties which, as we have discussed, can
naturally be explained by powerful radiatively driven winds.

In terms of individual source behaviour, which may differ from the overall trend due to individual characteristics e.g. precession, we can consider whether the spectral evolution with variability is consistent with our model and we do so on a source-by-source basis:\\

{\bf NGC 5408 X-1:} The {\it XMM-Newton} observations of this source cover a luminosity
range of only 7-9$\times$10$^{39}$ erg/s and show large and well
constrained fractional variability in each (Table 2). The X-ray spectrum changes
very little, with model parameters remaining consistent throughout and
similar fractional increases in soft and hard flux (from faintest to
brightest: Figure 5, Table 2). Although it would appear that the fractional
variability drops with increasing flux in the hard component, possibly
in keeping with our expectations of the wind tending towards homogeneity (see also Caballero-Garcia et
al. 2013), this is not well constrained. However, the spectral
and variability characteristics are fully consistent with a source that has been
viewed at moderate inclinations such that the wind is in our line-of-sight with large amounts of variability (dominated by Method 1).\\

{\bf NGC 6946 X-1:} There are clear spectral similarities between NGC 6946 X-1 and NGC
5408 X-1 (see Figure 5) with the source showing only
small total variations in flux (Table 2). We also detect large amounts of
variability; indeed NGC 6946 X-1 appears to demonstrate even larger
amounts of fractional variability than NGC 5408 X-1. It is therefore a
reasonable assertion that we have viewed the ULX at similar
inclinations. In addition to our analysis, NGC 6946 X-1 has been reported to have extreme (ultraluminous) UV emission ($\sim$4$\times$10$^{39}$ erg s$^{-1}$: Kaaret et al. 2010). Given the brightness this could be  associated with down-scattering in the wind and/or emission from the outer photosphere, and would imply moderate to high inclinations (depending on accretion rate - P07 - and any precession at the time).\\

{\bf NGC 5204 X-1:} The source appears to become considerably softer with increasing
luminosity (Figure 5) and the best-fitting model to the brightest
observation would imply that the hard emission becomes much
cooler (Table 2), although this is not well constrained. The evolution therefore appears consistent with
a line-of-sight changing such that the observer views increasingly into the
wind. As a result, the hard emission is progressively
beamed out of the line-of-sight with the remaining emission in the direction of the observer being increasingly Compton down-scattered. In our model we associate this behaviour with an inclination to the wind changing due to either precession or an increase in $\dot{m}_{\rm 0}$. \\

{\bf NGC 1313 X-1:} The evolution of this source (and X-2) has been the subject of many
papers (e.g. Pintore et al. 2012) and appears remarkably similar to
that of NGC 5204 X-1 (Figure 5) with a larger fractional increase in flux at
soft energies than hard, although it traverses a larger range in
luminosity (6-12$\times$10$^{39}$ erg s $^{-1}$). Once again we see a
spectrum (Obs.ID 0205230401) that appears to have a considerably cooler and
weaker hard component (Table 2). Although this is not the brightest observation, should we be viewing the source so that more of the wind enters the line-of-sight, then
a large portion of the intrinsic flux may be beamed out of the
line-of-sight and this may instead be at higher $\dot{m}_{\rm 0}$ or precessed.

{\it Unlike} NGC 5204 X-1, we can constrain the presence of
variability in two observations (Table 2). Both observations are at the
lowest flux levels (excluding Obs.ID 0205230401), consistent with our model where
the variability drops at higher luminosities (i.e. higher $\dot{m}_{\rm 0}$) in either Method.

Based on our model, we predict that our view of the ULX is evolving from low inclinations, where the wind does not substantially enter our line-of-sight so that the source appears hard with variability via Method 2, to one where the wind now enters our line-of-sight, the spectrum gets softer and variability becomes suppressed by increasing $\dot{m}_{\rm 0}$.\\ 

{\bf NGC 1313 X-2:} At the lowest fluxes, the source appears very similar to NGC 1313 X-1
and NGC 5204 X-1 (Figures 5 \& 6, Table 3) implying once again that we are
viewing at moderate inclinations to the wind. However, unlike the former sources, the
spectrum becomes increasingly hard with luminosity (Figure 6),
suggesting that the inner regions are not significantly obscured by the wind and are
instead geometrically beamed with increasing mass accretion rate, i.e. we are looking at small inclinations. 

At lower fluxes we detect variability (Table 3) which, given the
hard spectrum, we attribute to Method 2. We note that, once again the variability does not
appear in observations at higher luminosities even when the data
quality is extremely high, e.g. Obs.ID 0405090101. 

We also note the more recent observations presented by Bachetti et al. (2013) with two {\it XMM-Newton} observations of the source at low-moderate luminosities. Neither show constrained variability above 3mHz (our band of interest) but the brighter observation may show variability on longer timescales which could feasibly be explained by the source transiting along the softer branch of Figure 4 (or a combination of precession with decreasing $\dot{m}_{\rm 0}$).\\

{\bf HoII X-1:} The spectra generally appear very similar to those of NGC 5408 X-1 and
NGC 6946 X-1 (Figure 5 \& 6) although span a much greater range in
luminosity (3-10$\times$10$^{39}$ erg s $^{-1}$); we therefore also predict that we have observed the source at moderate inclinations.

We note the substantial difference in levels
of variability compared to NGC~5408~X-1 and NGC~6946~X-1 and, whilst we detect low
fractional rms in a single observation of HoII X-1
(Obs.ID 0112520601), this is poorly constrained. Certainly we do not detect variability in the highest
quality dataset (Obs.ID 0200470101) at the highest observed luminosity, which might imply that the wind has already tended towards
homogeneity, i.e. a higher $\dot{m}_{\rm 0}$. It is important to note that, in the case of HoII X-1, the detection of radio lobes (Cseh et al. 2014) suggest that the source is not likely to have been viewed at the smallest inclinations (the limit of not being strongly Doppler boosted gives a lower limit of 10 degrees). \\

{\bf HoIX X-1:} The evolution of HoIX X-1 has been studied by Vierdayanti et al. (2010) and
appears similar to NGC 1313 X-2 (Figure 6, Table 3). As the source is notable as one of the spectrally hardest ULXs, and appears to get harder with increasing luminosity, we predict that we are viewing at small inclinations to the wind so that it remains out of the line-of-sight throughout the existing observations. We attribute the general lack of variability (even given
the high data quality) to dampening due to high $\dot{m}_{\rm 0}$ (Method 2).

It is worth noting that Walton et al. (2014) observe spectral evolution of Ho IX X-1 across a broader energy bandpass using {\it NuSTAR}, in concert with {\it XMM-Newton} and {\it Suzaku}. The evolution may be consistent with a model of increasing hardness with luminosity (in the {\it XMM-Newton} bandpass) with the brightest epoch showing a cooler peak of the inner disc component, possibly due to increased Compton down-scattering in the (now more narrow) wind-cone. However, given present model degeneracies the exact nature of the evolution (in terms of spectral components) is not yet unambiguous.\\

{\bf IC~342 X-1:} The spectra of IC 342 X-1 and their evolution with luminosity appear
similar to those of Ho IX X-1 (Figure 6), implying inclinations that do not intercept the wind. In addition, we find that only the dimmest observation with adequate data quality (Obs.ID: 0206890201) shows
constrained variability (Table 3), which we associate with Method 2, implying that, as with HoIX X-1, the increase in brightness due to an increase in $\dot{m}_{\rm 0}$ damps the variability.\\ 

{\bf NGC 55 ULX-1:} The spectra of NGC 55 ULX-1 are the softest of our sample implying moderate to high inclinations to the wind. As variability is present, albeit at low levels, this implies a $\dot{m}_{\rm 0}$ low enough so that variability has not yet been fully damped. \\

Whilst the first order spectral-variability behaviour in individual sources as well as across the sample can be explained by our model, we can gain further insights by studying the nature of the energy dependence of the variability. Whilst past studies have used the fractional rms-spectrum (e.g. Middleton et al. 2011), here we extracted the covariance spectra (Wilkinson \& Uttley 2009) which have considerably smaller errors and allow a view of the correlated variability (relative to a reference band). In all cases, where statistics allow, we should expect the hard component to be variable due to obscurations (and potentially intrinsic disc emission should it not be thin) at larger inclinations, and an intrinsically variable hot inner disc (again, should it not be thin) and scatterings {\it into} the line-of-sight of the observer at smaller inclinations. The major difference we predict is that there could be some contribution to the shape of the covariance spectrum from down-scattering and changing $f_{\rm col}$ at small inclinations (although changes in $f_{\rm col}$ may also add variability to sources at larger inclinations, we expect the variability to be dominated by obscurations). In all cases we do indeed see that the hard component rather than the soft is varying with the disc normalisation tending to zero. This is potentially important as the covariance spectra of GX 339-4 and SWIFT J1753.5-0127 (Wilkinson \& Uttley 2009) when in the hard state (analogous to many ULXs should they contain more massive IMBHs), shows the disc to be variable with a similar proportion of disc to power-law as in the time-averaged spectrum (with the proportion of disc increasing to longer timescales). This is clearly not the case in the sources here (see the black residuals in Figures 9-11) as the disc contributes negligible variability, and implies that we are not seeing the expected behaviour for IMBHs ($>$ 100s of M$_{\odot}$). Unfortunately, the data quality is not presently high enough to confirm possible hints of down-scattering in the covariance spectra of those sources we have identified as viewed at small inclinations to the wind. However, based on the softening spectra, we predict that NGC 55 ULX-1 has been observed with the wind increasingly entering our line-of-sight by precession or increasing $\dot{m}_{\rm 0}$. The covariance spectra also show constrained evolution and imply that the variable component in the second observation has intercepted cooler/optically thicker material to that seen by the rest of the hard emission, inconsistent with models where the hard component is seen directly (e.g. disc-corona models of BHBs). 

The covariance spectra have proven valuable not only in allowing us to test consistency with predictions but also for highlighting the issues associated with spectral fitting which can often prove degenerate. In the case of our ULX sample we note that the time-averaged, best-fitting, high-energy spectral component ({\sc nthcomp}) overestimates the true component at soft energies. This results from tying the two spectral components together to ensure physical energy balance (section 3) but does not account for the changing temperature profile of the wind plasma. As a result, the soft component is not modelled accurately (by {\sc diskbb} in our fitting) and could conceivably allow for contribution from advection (P07). Whilst this does not cause problems for the inferred spectral evolution with variability (nor the apparent lack of a disc component in the covariance spectrum), it highlights the need for caution when obtaining characteristic temperatures and spectral parameters from fitting to broad continua. However, the variability spectrum should provide a means to model the soft component more accurately in future, thereby allowing us to probe its nature more thoroughly.  We note that, whilst the covariance spectra point towards issues in characterising the soft component, the rollover at high energies should be unaffected. Whilst the nature of this component is still unclear, if we identify it with the hot inner disc (see e.g. Walton et al. 2014) then the change in position of the rollover across the population should result from the changing amount of wind down-scattering the peak to lower energies. Whilst this seems broadly consistent with those sources where we believe the wind to have entered our line-of-sight (e.g. NGC 5204 X-1, NGC 1313 X-1, NGC 55 ULX-1: see Figures 5 \& 7), quantifying the change in temperature is a relatively complex issue but one we will address in a future work.

\section{Conclusions}

In this paper we have presented a model that results from considering the
structure of the super-critical accretion flow (Shakura \& Sunyaev
1973; P07; King 2009), clumps formed through radiative-hydrodynamic
instabilities (Takeuchi et al. 2014) and propagating mass accretion
rate fluctuations (Lyubarskii 1997; Arevalo \& Uttley 2006; Ingram \&
Done 2012). This invokes two methods of generating variability; via
clumps themselves on short timescales and via longer timescale trends
imprinted via propagating fluctuations. Combining the expected changes in the spectrum with the variability
leads to a set of spectral-timing predictions which depend on
inclination and accretion rate (section 3 and Figure 4). In order to test our predictions, we have analysed a sample of nine ULXs which are
bright in flux (thereby providing the highest-quality available
statistics) and avoid those ULXs which are likely to be BHBs with less extreme mass transfer rates. Although there is considerable scatter, we find that the trend of spectral hardness with variability power - taking the sample as a whole - is
consistent with the predicted evolution (Figures 4 \& 8) and a significantly negative slope which is inconsistent with accretion onto IMBHs (if we assume these would follow the ubiquitous path in hardness-variability of BHBs: Belloni 2010; Mu{\~n}oz-Darias et al. 2011). Such a finding is therefore consistent with the spectral properties being associated with super-critical accretion (as suggested by the mass measurement of a bright ULX: Motch et al. 2014).

We present the covariance spectra of those sources where data
allows for the first time. We note that these are broadly consistent with the expectation of obscuration/scattering/intrinsic variability of the inner disc (i.e. only the hard component varying), although the quality is not sufficient for detecting the subtle deviations we predict should be present. Whilst the shapes of the covariance spectra are also consistent with a variable high energy component/corona as seen in BHBs (although we strongly rule out a power-law shape for this component in the time-averaged spectra), we note that there appears to be very little/no contribution in any source by the soft component, inconsistent with what is seen in BHBs in the low-hard state (Wilkinson \& Uttley 2009), where disc reprocessing leads to a variable component. Whilst this appears to indicate that the 
 emission is not analogous to that seen in the low-state of BHBs, we also reiterate that, more importantly, the global trend in hardness-variability does not appear to match predictions for such a model invoking analogous accretion onto IMBHs.

Whilst the covariance spectra in general do not show evidence that unambiguously links the sources to the model we have presented, in the specific case of NGC 55 ULX-1 we observe evidence for down-scattering of the radiation, which argues against the variable hard component being seen directly. Finally, the covariance spectra highlight the present degeneracies
inherent in modelling ULX spectra; notably the contribution of the soft component was generally underestimated, although future studies making use of such a technique should be able to better determine the shape of the soft component and more accurately constrain its evolution with luminosity.

Although other models may be able to explain {\it some} of the behaviours of ULXs, the one we present here benefits from also being able to explain the lack of narrow atomic features in the
X-ray spectra of the hardest ULXs (e.g. Walton et al. 2012; 2013b) - explained by the inner, scattering
surface of the wind-cone being highly ionised by the strongly beamed
flux - and possible
presence of strong broad absorption features in the softest ULXs
(Middleton et al. 2014). We briefly note that variability caused by scattering from clumps in
the manner described in our model will lead to variable illumination
onto the optically thin expanded wind at larger radial distances
(directly and indirectly via reprocessing). Although we recognise that
there are multiple explanations, should the ionisation state of this
material change as a result, then we should expect changes in the
(possibly broad) atomic features (Middleton et al. 2014) which could
then explain the `soft lag' discovered in the cross spectrum of NGC 5408 X-1 (Heil \&
Vaughan 2010, De Marco et al. 2013). Another possible explanation is a change in the soft continuum in response to the variable hard emission (with the lag corresponding to the extra light travel time); the covariance spectra of NGC 5408 X-1 would imply that the soft component as a whole does not contribute significantly to the variability but we cannot rule out a ring of material with a small area being illuminated at the inner edge of the wind. Such behaviour may place independent constraints on the geometry and we will
investigate this in a future, dedicated paper.

There is clearly still a great deal of work left to be done in exploring the more complex predictions of such a model and any outliers that could be associated with unusual and exciting objects (e.g. HLX-1: Farrell et al. 2009, M82 X-1: Pasham, Strohmayer \& Mushotzky 2014, M82 X-2: Bachetti et a. 2014). Key to making progress will be to utilise the new powerful spectral-timing techniques, investigate the nature of the broad-band spectra (as {\it NuSTAR} is demonstrating to great effect) and obtain high quality spectra from multiple epochs, which next-generation missions such as {\it ATHENA} will regularly produce.

\section{Acknowledgements}

The authors thank the anonymous referee for helpful suggestions, and Tom Maccarone, Phil Uttley, Adam Ingram and Andrew King for useful discussion. MJM appreciates support via ERC grant 340442. DJW is supported by ORAU under the NASA Postdoctoral Program. TPR was funded as part of the
STFC consolidated grant ST/K000861/1. This work is based on observations obtained
with {\it XMM-Newton}, an ESA science mission with instruments and
contributions directly funded by ESA Member States and NASA.

\label{lastpage}

\end{document}